\documentclass[fleqn,usenatbib]{mnras}

\usepackage[T1]{fontenc}
\usepackage{ae,aecompl}

\usepackage{fontawesome}

\usepackage{graphicx}	
\graphicspath{{imgs/}}
\usepackage{amsmath}	
\usepackage{amssymb}	
\usepackage{xcolor}
\usepackage{color, colortbl}
\usepackage[utf8]{inputenc} 
\usepackage{ulem}

\newcommand*\diff{\mathop{}\!\mathrm{d}}

\newcommand{\Msun}{\text{M}_{\odot}}	
\newcommand{\fb}{FIREbox}    
\newcommand{\fbb}{FB30}    
\newcommand{\emberone}{\textsf{EMBER-1}}	
\newcommand{\embertwo}{\textsf{EMBER-2}}	

\usepackage{newtxtext,newtxmath}

\title[Emulating baryons across cosmic time]{EMBER-2: Emulating baryons from dark matter across cosmic time with deep modulation networks}

\author[Bernardini et al.]{Mauro Bernardini,$^{1}$\thanks{\href{mailto:mauro.bernardini@uzh.ch}{\url{mauro.bernardini@uzh.ch}}} 
Robert Feldmann,$^{1}$
Jindra Gensior,$^{2}$ 
Daniel Anglés-Alcázar,$^{3,4}$\newauthor
Luigi Bassini,$^{1}$ 
Rebekka Bieri,$^{1}$
Elia Cenci,$^{1}$
Lucas Tortora,$^{1}$
Claude-André Faucher-Giguère$^{5}$
\\
$^{1}$Center for Theoretical Astrophysics and Cosmology, Department of Astrophysics, University of Zurich,\\
$^{2}$Institute for Astronomy, University of Edinburgh, Royal Observatory, Blackford Hill, Edinburgh EH9 3HJ, UK, \\
$^{3}$Department of Physics, University of Connecticut, 196 Auditorium Road, U-3046, Storrs, CT 06269-3046, USA\\ 
$^{4}$Center for Computational Astrophysics, Flatiron Institute, 162 5th Ave, New York, NY 10010, USA\\
$^{5}$Center for Interdisciplinary Exploration and Research in Astrophysics (CIERA) and Department of Physics and Astronomy, Northwestern University,\\1800 Sherman Ave, Evanston, IL 60201, USA\\
}
\date{Accepted XXX. Received YYY; in original form ZZZ}

\pubyear{2022}

\begin{document}
\label{firstpage}
\pagerange{\pageref{firstpage}--\pageref{lastpage}}
\maketitle

\begin{abstract}
Galaxy formation is a complex problem that connects large scale cosmology with small scale astrophysics over cosmic timescales.
Hydrodynamical simulations are the most principled approach to model galaxy formation, but have large computational costs.
Recently, emulation techniques based on Convolutional Neural Networks (CNNs) have been proposed to predict baryonic properties directly from dark matter simulations.
The advantage of these emulators is their ability to capture relevant correlations, but at a fraction of the computational cost compared to simulations.
However, training basic CNNs over large redshift ranges is challenging, due to the increasing non-linear interplay between dark matter and baryons paired with the memory inefficiency of CNNs.
This work introduces \textsf{EMBER-2}, an improved version of the \textsf{EMBER} (\textbf{EM}ulating \textbf{B}aryonic \textbf{E}n\textbf{R}ichment) framework, to simultaneously emulate multiple baryon channels including gas density, velocity, temperature and HI density over a large redshift range, from $z=6$ to $z=0$.
\textsf{EMBER-2} incorporates a context-based styling network paired with Modulated Convolutions for fast, accurate and memory efficient emulation capable of interpolating the entire redshift range with a single CNN.
Although \textsf{EMBER-2} uses fewer than 1/6 the number of trainable parameters than the previous version, the model improves in every tested summary metric including gas mass conservation and cross-correlation coefficients.
The \textsf{EMBER-2} framework builds the foundation to produce mock catalogues of field level data and derived summary statistics that can directly be incorporated in future analysis pipelines.
We release the source code at the official website \href{https://maurbe.github.io/ember2/}{\faicon{github}}.
\end{abstract}




\section{Introduction}\label{sec:introduction}


Understanding the co-evolution of dark matter and baryons in the Universe is one of the central problems of galaxy formation.
The current theory predicts that galaxies form within dark matter halos from condensed gas that is accreted through the filamentary structure of the Cosmic Web.
Due to this direct interplay between dark and baryonic matter, the formation, growth and dynamical evolution of the galaxy and its parent halo are tightly connected with each other \citep[e.g.][]{Diemand2007}. 

While the gas distribution traces the underlying dark matter density on large scales that are still in the linear regime, the density distribution of the two fields increasingly decouples on smaller scales due to impact of baryonic physics on the matter distribution \citep[see e.g.][]{Borrow2020, Gebhardt2024}.
For instance, groups of galaxies can remove the gas supply of infalling galaxies through stripping mechanisms \citep[e.g.][]{Balsara1994, Masao2000, Bullock2001, Lokas2012, Barber2014, Wang2017, Lokas2020}.
In less massive haloes, cold gas streams can reach the galactic disk directly \citep[e.g.][]{Keres2005, Dekel2006, Brooks_2009, Faucher2011, Woods2014, vanDenVoort2015, Ho2019, Stern2020, Wang2022, Decataldo2024}, whereas for large haloes the gas is first shock-heated and requires longer to cool and settle onto the galactic disk \citep[e.g.][]{Birnboim_2003, Brooks_2009, Correa_2018, Stern2020, Faucher_2023}.
Thus, the gas accretion onto galaxies has a large dependence on halo mass and cosmic time \citep{Sanchez-Almeida2014, Wright2024}. In addition to environmental effects, internal processes such as star formation and feedback further shape the distribution of baryons in the Interstellar Medium (ISM) and strong feedback-driven outflows have been shown to affect the Circum-Galactic Medium (CGM) \citep[][]{Dubois2013, Angles2014, Choi2015, Faucher2015, Faucher2016, Angles2017, Hopkins2018, Li2018, Biernacki2018, Valentini2019, Hafen2019, Hafen2020, Donnari2021, Pandya2021, Xu2022, Wright2024}.

Cosmological hydrodynamical simulations are the most principled tool to study galaxy formation and its dependency on the environment.
The trade-off between box size and particle resolution is a key limiting factor for simulations. While simulations with higher resolutions and arguably more physically motivated ISM prescriptions are favored, they are restricted to smaller cosmological volumes \citep[see e.g. figure 2 in][and references therein]{Feldmann2023}.
Due to this trade-off, galaxy formation simulations rely on subgrid models as effective implementations of physical processes below the resolution scale of the simulation \citep[see e.g. the reviews by][and references therein]{Vogelsberger2020, Crain2023}. These subgrid models include prescriptions for gas cooling, star formation and feedback processes, and are a vital and active subject of research as they impact global galaxy features such as e.g. stellar masses \citep{Crain2015, Bottrell2017, Wellons2023}, morphologies \citep{Snyder2015, Dubois2016, Correa2017, Rodriguez-Gomez2019, Bluck2019, Wang2019, Gensior2024}, the ISM and CGM structure \citep{Bournaud2019, Roca2021, Bieri2023, Strawn2024}.\\

From these arguments, two aspects emerge:
i) The co-evolution of dark and baryonic matter on small scales strongly depends on the physics of galaxy formation, hence learning the mapping between dark and baryonic matter allows to quantify the effect of galaxy formation on small scales through its impact on the matter distribution.
ii) Large volume simulations targeting to model this co-evolution suffer from the volume resolution trade-off and largely depend on manually calibrated subgrid models. 
Higher resolution (and thus more realistic) simulations are needed to capture the physics more accurately, but are prohibitively expensive due to this constraint.\\




Deep Learning (DL) has been established as a complementary tool to alleviate exactly those problems for galaxy formation \citep[e.g.][]{Cohen2020, Villaescusa2020b, Villaescusa2021, Villaescusa2021b}. 
In particular, Neural Networks (NNs) have been used as super-resolution algorithms to increase the resolution of a given input field, such as surface density or temperature maps. Paired with lower resolution simulations this technique is capable of creating large high resolution mocks breaking the volume-resolution limitations \citep[e.g.][]{Li2021, Ramanah2020, Bernardini2022, Jamieson2023, Kaushal2022}. NNs have also been used to emulate cosmic fields from specific input variables (e.g. painting baryons onto dark matter or generating dark matter maps from specified cosmological parameters), with the advantage that emulation is orders of magnitudes faster than simulating the same fields \citep[e.g.][]{Wadekar2021, Troester2019, Zamudio2019, Thiele2020, Dai2021, Bernardini2022, Hassan2022, Lovell2022, Harrington2022, Horowitz2022}.

Many of the cited works use a special class of NNs, called Generative Adversarial Networks \citep[GANs,][]{Goodfellow2014, Goodfellow2016}. GANs are capable of learning mappings that exhibit a certain stochasticity by modelling the underlying probability distribution of the dataset. 
Many real-world examples are based on stochastic mappings, since many outputs are possible for a single input. 
In these scenarios, low-level supervised learning metrics (e.g. mean-squared error) fail as they are not designed to capture the probabilistic nature of the mapping.
The key advantage of GANs is that the high-level metric, capable of discriminating between real and fake data distributions, is directly learned in the training process.
Since galaxy formation is inherently a stochastic process \citep{Genel2019, Keller2019}, this aspect of GANs is particularly attractive.\\

In \cite{Bernardini2022} we introduced the \textsf{EMBER-1} framework to emulate high-resolution gas fields from dark matter inputs at a fixed redshift $z=2$. Predicting new gas realizations with \textsf{EMBER-1} is about $3\times 10^4$ times faster than the simulation runtime. However, it is limited to a single redshift while only emulating total gas and HI surface densities. 
In this work we extend the \textsf{EMBER-1} framework to train NN models that can consistently emulate multiple baryon fields over a large domain of cosmic time, i.e. from $z=6$ to $z=0$.
This aspects poses a challenge on the modelling side, since most NN models are not designed to learn over large time-series. In the following paragraphs we describe the computational challenges associated with these extensions. The solutions presented in later sections are the main contribution of this paper.
\\

Cosmological structure and galaxy formation are not time-invariant problems, i.e. the growth of structure is dependent on redshift. 
The evolution of a galaxy in its environment can be seen as a time-series, where each snapshot encodes field aspects such as the dark matter distribution or the temperature-phase information of the gas component.
The environmental landscape changes drastically over the lifetime of a galaxy, due to the non-linear behavior of gravity and astrophysics.
As a consequence, dark matter and baryons increasingly decorrelate over cosmic time (see figure \ref{fig:cross_corr}). This introduces a major problem when modelling their relation with NNs, since the mappings exhibit a strong redshift dependence. Thus, a NN trained on a single redshift slice (e.g. $z=2$ for \textsf{EMBER-1}), is almost guaranteed to fail at different redshifts, since the learned correlations do not adapt accordingly.
In principle, it is possible to increase the number of trainable model parameters to adapt to multiple redshifts, however such models suffer from severe memory bottlenecks and might not fit on modern GPUs \citep{Hascoet2019}. 
Compared to the EMBER-1 model presented in \cite{Bernardini2022} that was fitted to predict total gas and HI density, a more complete field emulator must include an efficient prescription for learning correlations across the entire redshift space in a modular fashion.\\

\begin{figure*}
    \includegraphics[width=\textwidth]{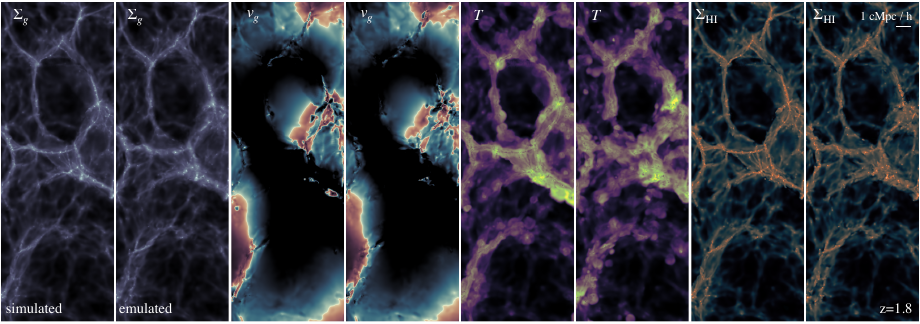}
    \caption{
    Overview figure showcasing simulated and emulated baryon fields at $z=1.8$.
    Each pair of panels shows the simulated and emulated fields from the \fbb{} simulation and the \embertwo{} model respectively using the same color scale.
    The direct comparison showcases the ability of the model to produce realistic predictions on the field level for the gas and HI surface densities, radial gas velocities and temperatures.
    A video version across cosmic time of the above comparison can be found at the official \embertwo{} \href{https://maurbe.github.io/ember2/\#movies}{website}.
    }
    \label{fig:overview}
\end{figure*}
In this work, we build on the results presented in \cite{Bernardini2022} and expand the models to address the problem of emulating baryon fields with NNs across cosmic time. In figure \ref{fig:overview} we show an overview of our method and its objectives.
The paper is structured as follows. In section \ref{sec:simulations} we briefly introduce the simulations and describe the pipeline to produce the datasets used in this work. In section \ref{sec:ember} we explain the mappings, discuss the architecture and the training routine for the DL model. The results are presented and discussed in section \ref{sec:results}. Finally, we propose future applications of our method and conclude in section \ref{sec:conclusion}.

\section{Simulations and data generation}\label{sec:simulations}
To create our dataset for the machine learning model, we use a simulation that is part of the Feedback in Realistic Environments (\textsc{FIRE}\footnote{See the official \textsc{FIRE} project website: \url{https://fire.northwestern.edu}}) project. 
Similar to the approach in \cite{Bernardini2022} we use a cosmological hydrodynamical simulation to create our dataset.
This simulations is run with {\sc gizmo} \citep{Hopkins2015}\footnote{A public version of \textsc{GIZMO} is available at \url{http://www.tapir.caltech.edu/~phopkins/Site/GIZMO.html}}, which implements a multi-method gravity and hydrodynamics solver. Furthermore, the simulations use the FIRE-2 physics model \citep[][]{Hopkins2018}.
The multi-scale initial conditions tool MUSIC \citep[][]{Hahn2011} is used to create the initial conditions for the simulations, which are run with the Planck 2015 cosmology \citep[][]{Planck2016}: $H_0$= 67.74 km/s/Mpc, $\Omega_m$=$0.3089$, $\Omega_\Lambda$=$0.6911$, $\Omega_b$=$0.0486$, $\sigma_8$=$0.8159$ and $n_s$=$0.9667$. 
The FIRE-2 galaxy formation model has been validated in zoom-in and cosmological box simulations, and is in good agreement with observations with respect to scaling relations \citep[e.g.][]{Rohr2022, Feldmann2023, Bassini2024, Cenci2024a, Gensior2024}, realistic galaxy discs \citep[e.g.][]{Gensior2023, Cenci2024b} and large-scale HI properties \citep[e.g.][]{Bernardini2022, Tortora2023}.
In the following we briefly summarize the most important aspects of the simulation for this work and refer the interested reader to the corresponding references for further details.


The simulation used in this work is a high resolution hydrodynamical cosmological volume simulation with box length of $30\,\text{cMpc}/h$.
The physics of this run are identical to the \fb{} simulation presented in \cite{Feldmann2023}, although in a $8\times$ larger box and at reduced resolution. Hereafter we will refer to this simulation as \fbb{}.
The simulation is initialized with $512^3$ dark matter and $512^3$ gas particles with mass resolutions of $m_{\text{dm}}=7.86\times 10^6 \, \Msun/h$ and $m_{\text{gas}}=2.71\times 10^6 \, \Msun/h$. For dark matter particles the softening length is fixed to $320\,\text{pc}$ and gas particles have a variable softening lengths with a minimum of $16\,\text{pc}$.
Compared to the \fb{} simulation, \fbb{} extends the mass range of dark matter haloes from $7.8 \times 10^{12}\, \Msun /h$ to $3.8 \times 10^{13}\, \Msun /h$, as it contains more massive objects.
Since the scope of this work is primarily geared towards modeling cosmic gas in and around high mass dark matter haloes, we chose the \fbb{} simulation for the creation of our dataset. Moreover, due to the larger volume, intermediate scales in the simulation are less affected by cosmic variance, while more data samples can be generated, rendering the dataset more representative. Even though the resolution of the \fbb{} simulation is low compared to other FIRE simulation suites, it is still useful for training and validating neural network models.
\\

The task of our model is to learn the mapping from 2-dimensional dark matter fields including the projected dark matter surface density and their radial velocities to 2-dimensional gas surface density, gas radial velocities, gas temperature and HI surface density for the redshift interval $z=6$ to $z=0$, i.e. for each redshift the model learns the following probabilistic mapping
\begin{equation}
    f: (\Sigma_d, v_d) \rightarrow (\Sigma_g, v_g, T, \Sigma_\mathrm{HI}).
\end{equation}
We emphasize that the dark matter field used as input is extracted from the full hydro run. In principle the dark matter from the hydro differs from a pure dark matter simulation due to the presence of baryons \citep[see e.g.][]{Schneider2015, Arico2021, Sharma2024}.
In \cite{Bernardini2022} we have investigated the effects of this difference on the emulated matter power spectrum at $z=2$ and found that the predictions are similar when emulated from a hydro or a pure dark-matter only run.
However, it is a priori unclear whether this holds across a wide range of redshifts as investigated here. To address this question, we also use the dark-matter only counterpart of the \fbb{} simulation but only during the validation phase.

To train and test the algorithm we extract a total of 37 redshift slices from the simulation, where we use an equidistant spacing of 0.25 between redshifts $6 \geq z \geq 2$ and a spacing of 0.1 between $z \geq 2 \geq 0$.
As a first step we partition the simulation domain of the \fb{} volume into 39 equal slabs with thickness $1.5\,\text{Mpc}/h$ with an overlap of $0.75\,\text{Mpc}/h$ between two subsequent slabs. 
This partitioning is done for all three spatial dimensions, resulting in a total number of 117 slabs (for a specific redshift).
Next, similar to \cite{Bernardini2022}, we use \texttt{smooth} and \texttt{tipgrid} to project and deposit the particles onto 2-dimensional grids.\footnote{Code is publicly available at \url{https://github.com/N-BodyShop/smooth}.} 
Due to the large dataset, the grid resolution is chosen to be $\sim$29$\,\text{ckpc}/h$, such that a single slice contains $1024^2$ pixels. An example result of this procedure is shown in figure \ref{fig:overview}.
To build our training and testing datasets, we use a 3-fold cross-validation approach, where we use the projections from two axes for training while the remaining axis is used for testing.
In order to validate the models predictive accuracy, we also test the model on an independent simulation consisting of a 15 cMpc$/h$ volume with the same physics and resolution as \fbb{}. These results are presented in appendix \ref{app:test_independent_simulation}.


\section{\textsf{EMBER-2}}\label{sec:ember}
\subsection{Fitting multiple redshifts across resolutions}\label{subsec:challenges}
\textsf{\textsf{EMBER-2}} (\textbf{EM}ulating \textbf{B}aryonic \textbf{E}n\textbf{R}ichment) is the improved version of the \textsf{\textsf{EMBER-1}} framework introduced in \cite{Bernardini2022}.
In this section we highlight the central shortcomings of \textsf{EMBER-1} and propose solutions that we implement in \textsf{\textsf{EMBER-2}}.
The objective of \textsf{\textsf{EMBER-2}} is to train a generative model capable of synthesizing 2-dimensional gas fields from 2-dimensional dark matter information for the entire redshift range between $z=6$ and $z=0$.

The basic convolution operation, used in every convolutional neural network (CNN), operates on a fixed pixel scale and cannot account for the time-variance of the problem at hand. 
A great advantage of CNNs is that they are capable of identifying similar features regardless of their position in the image domain \citep{Cohen_2016}. 
Since high density regions are scarce in large volume simulations, learning features associated with these structures is not guaranteed to be successful. Furthermore, if we want to learn the mapping across multiple redshifts with a single NN, then we expect the required number of trainable parameters to scale linearly with the number of redshifts in the dataset, because the feature maps learned at $z=6$ might only be partially usable at $z=0$. Since our dataset contains 37 redshifts, approaches based on vanilla CNNs are thus challenging. At this point we emphasize that \textsf{\textsf{EMBER-1}} is based on vanilla CNNs and is therefore not a good solution to this problem.\

\subsubsection*{Modulated Convolutions}
\begin{figure}
    \includegraphics[width=\columnwidth]{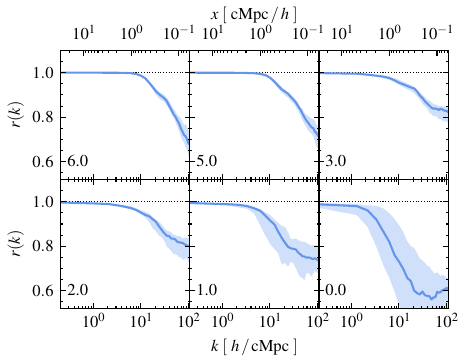}
    \caption{
    Cross-correlation coefficients $r(k)$ between simulated dark matter and total gas density in the \fbb{} simulation. $r(k)$ measures how decoupled the fields are at different scales and redshifts (bottom left corners).
    With decreasing redshift the decoupling scale shifts from $\sim$1 cMpc$/h$ at $z=6$ to $\sim$5 cMpc$/h$ at $z=0$.
    }
    \label{fig:cross_corr}
\end{figure}
In the following, we present an improved approach to solve the aforementioned problem. The continuous map $f$ between dark matter ($x=(\Sigma_d, v_d)$) and gas ($y=(\Sigma_g, v_g, T, \Sigma_\mathrm{HI})$) exhibits a redshift ($z$) dependency:
\begin{equation}
    y = f(x, z).
\end{equation}
The dark matter and total gas fields are strongly correlated at early times, whereas the fields decouple from one another with decreasing redshift. 
The cross-correlation coefficient is a physical measure that indicates how correlated two fields are. It is defined through the power spectra of dark matter ($P_d$) and gas ($P_g$) density and their corresponding cross-power spectrum $P_{d,g}^{\times}$, i.e.
\begin{equation}\label{eq:r}
    r(k) = \frac{P_{d,g}^{\times}}{\sqrt{P_d P_g}}.
\end{equation}
If $r(k)=1$ then the two fields are perfectly correlated at scale $k$.
In figure \ref{fig:cross_corr} we show $r(k)$ for different redshifts in the \fbb{} simulation. With decreasing redshift the scale of decoupling shifts to smaller $k$, due to feedback from galaxies that impact the matter distribution at Mpc scales. This behavior indicates that the information about the spatial distribution of gas that is stored in the dark matter, decreases with redshift.
Thus, we expect that learning the mapping $f$ at higher redshifts to be substantially easier compared to later times.

To further simplify the problem, we define $f$ such that it can be written with two functions $g$ and $w$ that split the dependency of the input variables, i.e.
\begin{equation}\label{eq:f}
    f(x, z) = g(x) \cdot w(z),
\end{equation}
where in practice, both $g$ and $w$ are represented by NNs. The idea behind this representation is that $x$ can be understood as a local variable whereas $z$ describes the global context independent of $x$ as it is shared throughout all samples $x$ at fixed redshift. Hence, at a fixed $z$, the mapping $f$ can only depend on the local dark matter field $x$, since the global context is factored out.\\

In order to build a memory efficient emulator learning across redshifts, we follow the Modulated Convolution (ModConv) approach first introduced by \cite{Karras2018}. We factorize the convolution kernels of the NNs rendering them adaptive to the global context information.
Similar approaches have been used by \cite{Horowitz2022}, \cite{Jamieson2023} and \cite{Zhang2024}.
ModConvs modulate each base convolution kernel with an external style vector $w$, which is learned from a set of input variables, in our case the redshift input $z$, i.e. $w=w(z)$.
This change in architecture extends the power of a single convolution kernel, as it can be modulated to operate on a variety of different redshifts. 
In the following we will describe the individual architecture components of \textsf{EMBER-2} representing the function $f$.

\subsection{Network architecture}
\begin{figure*}
    \includegraphics[width=\textwidth]{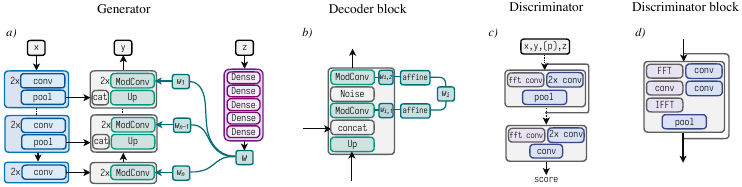}
    \caption{
    Network architectures and individual building blocks thereof.
    \textbf{\textit{a)}} Schematic overview of the generator $G$ based on the typical U-Net architecture. Features from the input $x$ are extracted via an encoder (blue blocks) comprised of two normal convolutions and a strided convolution for pooling. Subsequently the representations are decoded through a series of ModConv blocks to synthesize $y$. The Mapping network shown in purple is a simple 5-layer multi-layer perceptron that maps the global redshift information $z$ to the style vector $w$.
    \textbf{\textit{b)}} Zoom-in of a single block in the decoder branch of the synthesis network comprised of a bicubic upsampling step, followed by a concatenation with the skip connection from the corresponding encoder block. The base kernels of the two ModConvs are modulated via an affine layer with the externally mapped style vector $w$. Between the two convolutions we concatenate 8 channels of gaussian noise to the tensor before feeding it through the second ModConv. We emphasize that the two affine mappings, $w_{i,1}$ and $w_{i,2}$, are learned individually.
    \textbf{\textit{c)}} Schematic of the discriminator $D$ comprised of individual blocks featuring a two-path strategy of convolutions in real and Fourier space. The last convolution outputs a single scalar representing the score of the discriminator.
    \textbf{\textit{d)}} Discriminator block comprised of two normal convolutions on the right path. Along the left path the data is first transformed to Fourier space, convolved to extract frequency features and subsequently transformed back to real space via the inverse Fourier transform. The data from the left and right path are added element-wise and pooled by a strided convolution. Throughout all networks, we use kernel sizes of 3 and strides of 1 (except for the strided convolution, where a stride of 2 is used). Source code is available at the official \href{https://maurbe.github.io/ember2/}{github \faicon{github}} repository.
    }
    \label{fig:nn}
\end{figure*}

\subsection*{Generative Adversarial Networks}
Building upon the neural network model introduced in \cite{Bernardini2022}, we use conditional GANs (cGANs) as our model framework. 
Here we primarily discuss the improved building blocks and refer the interested reader to \cite{Bernardini2022} for a more detailed discussion of GANs.
Briefly, cGANs are comprised of two sub-networks called the generator $G$ and discriminator $D$ that compete in an adversarial game. While $G$ is trained to generate samples close to the true data distribution, $D$ is trained to distinguish between real and fake samples, corresponding to a min-max algorithm where both neural networks try to outperform their corresponding opponent. In terms of loss functions $\mathcal{L}$ for $G$ and $D$, the min-max game can be written as
\begin{align}
\label{eq:L_D}
\mathcal{L}_D &= +\mathbb{E}_{\eta} \left[ D(G(\eta|x)|x)) \right] \; - \; \mathbb{E}_{y} \left[ D(y|x) \right ] , \\
\label{eq:L_G}
\mathcal{L}_G &= -\mathbb{E}_{\eta} \left[ D(G(\eta|x)|x)) \right]
\end{align}
where $y$ and $x$ are samples from the true data distributions $p_y$ and $p_x$ and $\eta$ is the noise variable drawn from a normal distribution. $\mathbb{E}$ represents the expectation operator.

In the following, we will give a detailed overview of the exact architectures for $G$ and $D$. Individual important building blocks are shown in figure \ref{fig:nn} as a visual help for the following sections.

\subsection*{Generator (\href{https://github.com/maurbe/ember2/blob/91bf1b94de068eceec269fdee6b4e326e8e1f671/nn/modules.py\#L10}{code})}
Similar to \cite{Bernardini2022}, we implement our $G$ network as a classical encoder-decoder network, where an input $x$ is mapped to an embedding space that is reduced in spatial size and from which the embedded information is decoded to an output $y$ of the same spatial dimensions as $x$.
For this we use the U-Net architecture first introduced by \cite{Ronneberger2015}. The U-Net is designed with two branches, an encoding and a decoding part. As in the typical image-based autoencoder case, the information extraction and compression is achieved by means of convolutions followed by information pooling. In our case, we design an individual encoder block as two convolutions followed by a pooling operation for which we use strided convolutions (see panel \textit{a)} in figure \ref{fig:nn}). No pooling is used in the bottleneck (lowest) block of the encoder. Before the strided convolution, the data is copied along a skip connection that injects the learned embeddings into the corresponding decoder blocks. In \cite{Bernardini2022} we followed the original implementation by \cite{Ronneberger2015} where the number of feature channels is doubled for every subsequent convolution block. This results in a large number of feature channels especially in the bottleneck part, where the spatial data representations are small due to the repeated use of pooling operations. We empirically found that the encoder network performs equally well when using a fixed number of 64 channels in all convolution kernels in the encoder network. In this manner, the number of parameters is drastically reduced while training becomes faster with close to no performance loss.\\

The decoder network for our U-Net is substantially different from the one used in \cite{Bernardini2022} where we used basic convolutions rather than ModConvs. Compared to simple convolutions, ModConvs allow their kernel parameters to be modulated according to an external feature vector that factors in global information such as the redshift. The decoder network is based on the synthesis and mapping networks of StyleGAN2 \citep[][]{Karras2018}, but we introduce a few important changes to the architecture adapted to our problem. 
The mapping network, shown in purple in panel \textit{a)} of figure \ref{fig:nn}, is trained to map the redshift information $z$ to a style representation $w$ that is used to modulate the kernel parameters.
We design the mapping network as a simple multi-layer perceptron (MLP) with 5 layers and 64 units each, so that $w$ is a 64-dimensional vector. We found that this setup worked best for our approach.

A single block in the decoder network is comprised of the following operations. For all blocks, except the bottleneck, the data coming from the preceding decoder block is first upsampled by bilinear interpolation and subsequently concatenated with the embeddings from the skip connections from the encoder. Next, the style vector $w$ is mapped by two single dense linear layers (termed affine transformations $w_{i,1}$ and $w_{i,2}$) and multiplied with the corresponding kernel parameters of the first and second ModConv. The data is then convolved by the modulated convolution kernels. After the first ModConv we concatenate 8 channels of random noise drawn from a normal distribution to account for the stochastic nature of the mapping between $x$ and $y$. Through this noise injection the generator can learn to synthesize features that are not present in the input $x$.
The reason for using a fixed number of noise channels for all decoder levels is two-fold. Firstly, it massively reduces the memory footprint of the computation.
Second, we found that concatenating the noise instead of element-wise addition works better for our approach, since the network can decide how much noise to inject at the different network levels through the second ModConv. A schematic depiction of the computation steps is shown in panel \textit{b)} of figure \ref{fig:nn}.

\subsection*{Discriminator (\href{https://github.com/maurbe/ember2/blob/91bf1b94de068eceec269fdee6b4e326e8e1f671/nn/modules.py\#L89}{code})}
The discriminator follows the classical architecture found in many image-based GANs \citep[e.g.][]{Heusel2017, Gulrajani2017, Arjovsky2017a, Arjovsky2017b, Karras2017, Karras2018, Karnewar2019}.
The general architecture is composed of individual blocks that operate at different scales through a combination of convolutions and pooling operations (see \textit{c)} of figure \ref{fig:nn}) similar to the Generator architecture. 
Recently, many works have pointed out that standard GANs based on spatial convolutions fail to learn the high frequency features in images. This phenomenon, termed spectral bias, is present in all CNN models that are based on spatial convolutions. In particular, standard discriminators in GANs are not capable of distinguishing between real and fake samples, even though the difference in Fourier space is remarkable \citep[e.g.][]{Dzanic2019, Khayatkhoei2020, Jung2020, Chen2020, Fuoli2021, Mao2021, Wang2021, Schwarz2021, Chandrasegaran2021, Zhou2022}.
Many works have investigated different strategies to combat this shortcoming. Broadly speaking, the proposed solutions are either modifying the loss functions by rendering them more sensitive to the frequency discrepancies \citep[e.g.][]{Fuoli2021} or introducing changes in the architecture of both $G$ and $D$ to learn spectral features \citep[e.g.][]{Jung2020, Wang2021, Chen2020, Mao2021}.

For our model we follow a similar approach to \cite{Mao2021}, who proposed the Residual Fourier Unit (RFU) to learn embeddings directly in Fourier space and thereby improve the spectral bias of GANs. 
We use Fourier features in the design of our discriminator blocks, which are depicted in panel \textit{d)} of figure \ref{fig:nn}.
In particular, a single block is composed of two branches. Along the first branch the data is first transformed to Fourier space by means of the Fast Fourier Transformation (FFT), subsequently convolved and transformed back to real space by the inverse FFT (IFFT). The second branch is composed of two spatial convolutions and an element-wise addition with the data from the Fourier branch, followed by a pooling operation implemented as a strided convolution. Additionally, we add the input data to the output yielding a residual connection that counteracts vanishing gradients \citep[][]{He2015}. In this manner, we allow the discriminator to extract and merge information from the spectral and spatial domain simultaneously without modifying any loss functions.\footnote{We also tested RFUs in the generator architecture and found no improvement. Since the use of RFUs results in slower training times, we do not use them for $G$.}
Throughout both networks $G$ and $D$ we use LeakyReLU activation functions \citep{Maas2013}.

A second problem that is limited to cGANs is that they tend to ignore the injected noise during training and solely focus on the conditional input variable. 
This behavior can be seen as a form of mode collapse where $G$ does not learn to interpolate the whole data distribution but merely learns it at the specific data points in the dataset by ignoring the noise contribution \citep[see e.g.][and references therein]{Mathieu2015, Isola2016, Zhu2017, Yang2019}. 
Many works have proposed different strategies to combat this behavior by e.g. modifying the loss function of the discriminator \citep[][]{Zhu2017, Yang2019}.
Here, we follow a different approach, similar to \cite{Adler2018}, that is focused on an architectural change in the discriminator.
The key difference to the conventional GAN training is that during a single iteration two samples $p_1$ and $p_2$ are generated with different noise realizations $\eta_1$ and $\eta_2$. Together with the conditional input variable $x$ and the true variable $y$, we build a \textit{real} tuple $(x, y, p_2)$ and a \textit{fake} tuple $(x, p_1, p_2)$ with the corresponding labels. Both vectors are subsequently used to train the discriminator, where the loss function now has the form 
\begin{equation}\label{eq:L_D}
\mathcal{L}_D = +\mathbb{E}_{\eta_1, \eta_2} \left[ D(p_1, p_2|x)) \right] \; - \; \mathbb{E}_{y, \eta_2} \left[ D(y, p_2|x) \right ].
\end{equation}
If $G$ now ignores the noise variables, then $p_1\approx p_2$ and the discrimination for $D$ becomes trivial. Hence, the feedback from $D$ forces $G$ to include the noise variable in the synthesis process to learn the distribution of possible realizations, such that $p_1 \not \approx p_2$.

\subsection*{Training}
For the training process we prepare individual data samples in the following way. As described in section \ref{sec:simulations}, we deposit the gas particles density onto a uniform grid with pixel resolution of 29 ckpc$/h$.
Training NNs is greatly improved when the data values are of $\mathcal{O}(1)$. 
As pointed out in many works before \citep[see e.g.][]{Wadekar2021, Thiele2020, Bernardini2022} the choice of the data normalization scheme plays an important role for the task of predicting cosmic fields with correct statistics. Depending on the cumulative distribution function we scale the data samples with the following transformations:

\begin{align}\label{eq:scaling}
    &\Sigma_d \rightarrow \sinh \left[\frac{1}{4} \lg \left(10^{-4} \, \Sigma_d / [\Sigma] + 1\right) \right] &\text{dm density}&\\
    &\Sigma_g \rightarrow \sinh \left[\frac{1}{4} \lg \left(10^{-3} \, \Sigma_g /  [\Sigma] + 1\right) \right] &\text{gas density}&\\
    &T \rightarrow \sinh \left[ \lg \left(10^{-3} \, T / \mathrm{K} + 1\right) / m_t(z) \right] &\text{gas temperature}&\\
    &\Sigma_\mathrm{HI} \rightarrow S\left(\lg \left(10^2 \Sigma_\mathrm{HI} / [\Sigma] + 1\right) \right) &\text{HI density}&\\
    &v_d \rightarrow s\left(\frac{v_d}{\mathrm{km \, s^{-1}}}\right) &\text{dm radial velocity}&\\
    &v_g \rightarrow s\left(\frac{v_g}{\mathrm{km \, s^{-1}}}\right) &\text{gas radial velocity}&
\end{align}
where $s(x)$ is a symmetric logarithmic\footnote{We use $\lg$ representing $\log_{10}$ throughout this paper.} function and $[\Sigma]= (\Msun / \rm ckpc^2)$.
The temperature fields show a large redshift dependence, and thus we use a redshift-dependent normalization factor $m_t(z)=-0.193 \times z + 5.346$. The exact values have been chosen, such that the maximum values of the transformed temperatures are $\mathcal{O}(1)$. Since the distribution of the physical HI field has a very extended tail of high density values, we use a methodology termed histogram equalization. 
Histogram equalization can transform a sample distribution into a target distribution by shifting individual sample values.
In practice this can easily be achieved by ordering the sample values and then modify the values such that the cumulative distribution between the sample and target match in each sample bin.
In our case, we choose the dark matter density field as the equalization target for the HI fields. We then fit $S(x)$ and obtain an analytical expression for the transformation of HI density field. The analytic expression is important, since we also need the inverse of the transformation for our pipeline.
For some of the fields, we use an additional $\sinh$ function to stretch the range of high density pixels, since many summary statistics such as spectral metrics greatly depend on the values in the high density regime (concrete examples are presented in section \ref{sec:results}).\\

According to the $k$-fold split of our dataset with $k=3$, we train 3 models. During training, we randomly crop images of size $64^2$ corresponding to a comoving size of 1.875 cMpc$/h$.
This particular choice was primarily motivated due to the large memory consumption and longer training times when training on larger patches.
We emphasize that introducing a fixed pixel resolution restricts the model to this specific resolution scale, i.e. the resolution between training and application must remain the same.
However, due to the fully convolutional design of the generator network, entire 30 cMpc$/h$ slices can be emulated directly during prediction.

The \emberone{} model we presented in \cite{Bernardini2022} was only capable of separately emulating total gas and HI density at a fixed redshift of $z=2$. For \embertwo{} we relax these constraints and emulate multiple baryon channels over a large domain of cosmic time.
To train \embertwo{} we use the objectives presented in equations \ref{eq:L_G} and \ref{eq:L_D}. Different to the previous training strategy in \cite{Bernardini2022}, we do not use additional pixel-based losses and revert to a purely adversarial loss. 
\embertwo{} is massively reduced in size in terms of trainable parameters. The generator consists only of 4.5$\times 10^6$ parameters, i.e. 6$\times$ smaller than \emberone{}.
The architectural improvements in \embertwo{}, the slimmed encoder network and especially the ModConv scheme in the decoder of $G$, allow for more efficient training with less parameters, feasible memory consumption and faster inference times while operating on larger datasets.
As a result, an entire \fbb{} mock catalog can be emulated on a single GPU within $\sim$5 seconds.
We present additional information regarding implementations and training of the model in appendix \ref{app:additional_information}.

\section{Results and Discussion}\label{sec:results}
In this section we discuss the results obtained from training the NN model to solve the problem of emulating gas fields. Since the emulation is not bound to any physical constraints it is crucial to check whether the emulated physics are consistently captured by the model. The realness in a physical context of the samples of $G$ is only constrained through the discriminative power of $D$, i.e. no physics-based loss functions are used. 
To this end we investigate key concepts such as mass conservation and higher order moments such as cross-correlations between dark matter and gas fields. Regarding the physical realness of samples, we furthermore investigate phase-space density reconstruction and temperature-density phase diagrams.
For the following sections we provide additional video material as visual aids at the official \textsf{EMBER-2}  \href{https://maurbe.github.io/ember2/}{website}.\\

\subsection{Probing the distribution of total gas surface density}
\begin{figure*}
    \includegraphics[width=\textwidth]{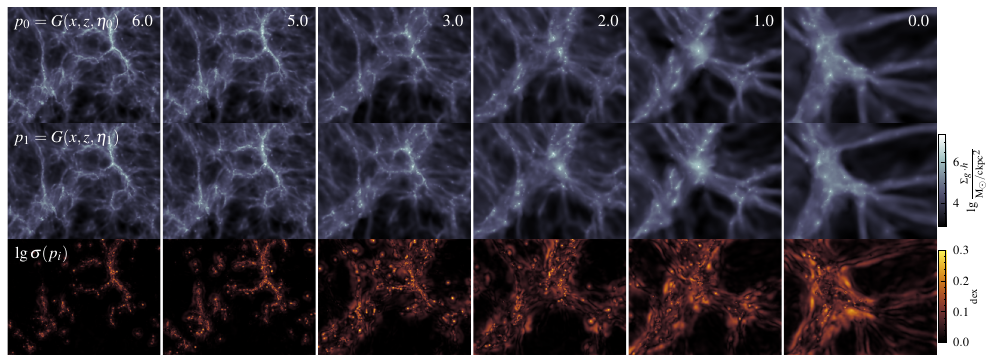}
    \centering
    \caption{
    Summary grid showcasing the performance of \embertwo{} across the trained redshift range. The images show a region of 5.5 cMpc$/h$ $\times$ 7.5 cMpc$/h$. The first two rows show example patches of evolving gas density at different redshifts (indicated in the top-right corner) of the same cosmic region but generated with two distinct noise realizations $\eta_1$ and $\eta_2$. 
    In the third row we smoothly vary the noise variable $\eta$ to produce 128 different noise realizations. From these predictions we show the per-pixel standard deviation, where 
    brighter regions indicate larger standard deviations and thus regions that are more subject to large changes between different realizations. We also see how the noise impacts increasingly larger scales for decreasing redshift. This behavior indicates a weaker cross-correlation between the dark matter and gas densities for decreasing redshifts, which is in agreement with the simulation (see figure \ref{fig:cross_corr}).
    }
    \label{fig:base_visuals}
\end{figure*}

The generative part of the model is trained to learn the distribution parameterized by three input features: the conditional dark matter fields $x=(\Sigma_d, v_d)$, the global context variable consisting of the redshift $z$ and the noise input $\eta$. Sampling this distribution allows to investigate the internal scatter the model has learned. More concretely, the amount of emulated CGM and IGM halo gas will vary depending on the exact environmental dark matter distribution and the noise vector $\eta$. To test how the three input variables $x, z$ and $\eta$ alter the outcome of the model prediction, we vary one or more and fix the remaining variables. The results shown in figure \ref{fig:base_visuals} should act as a visual aid.\\
\newline 
\textbf{Varying $x$ and $z$ (\href{https://maurbe.github.io/ember2/\#movies}{video}):} In the first row of figure \ref{fig:base_visuals} we show the emulated $\Sigma_g$ across different redshifts ($z=6$ to $z=0$). 
These regions were emulated by feeding a dark matter patch $x$ at redshift $z$ to the model while fixing the noise input $\eta$. 
The visual summary clearly showcases the models capabilities of generating realistic samples across different redshifts. Furthermore, the model has learned to encode the additional noise input resulting in very diverse small scale structures as well as larger features on Mpc scales.
This test highlights the capability of the ModConv scheme to capture the relevant modes in the input variable $x$ and additionally generate small scale features that showcase a strong redshift dependence.
We emphasize that varying $x$ and $z$ simultaneously is the basic mode of operation of the \embertwo{} model.\\
\newline
\textbf{Varying $\eta$ (\href{https://maurbe.github.io/ember2/\#movies}{video}):} Especially at lower redshifts, the model shows clear behavior of successfully generating gas features on Mpc scales. These features stem from the injected noise, which is especially interesting when recalling that with decreasing redshift the input and target fields become more decoupled (as shown in figure \ref{fig:cross_corr}).
As a visual quantification of this behavior, for each redshift $z=6, 5, 3, 2, 1$ we fix the input variable $x$ and smoothly interpolate the noise parameter $\eta$ to understand which scales are primarily affected by the noise injection. 
Specifically, for each redshift and slice $x$ we produce 128 samples with different noises, s.t. we end up with 128 boxes for testing.
We show two distinct samples $p_0$ and $p_1$ in the first two rows of the summary grid in figure \ref{fig:base_visuals} corresponding to two different noise realizations $\eta_0$ and $\eta_1$.
We subsequently stack and reduce the 128 samples by computing the standard deviation per-pixel. This results in the map shown in the third row of figure \ref{fig:base_visuals}, where brighter regions indicate a higher variability. 
Clearly, at earlier times, primarily the high density peaks and their close environment are affected since the fields (and thus the mapping) are still in the linear regime of structure growth. Similarly, the high density peaks are also subject to change at lower $z$, because the NN learns to reproduce the intrinsic scatter of the gas mass to halo mass relation as shown in \cite{Bernardini2022}.
For lower redshifts the model has successfully learned that regions between haloes and large-scale regions in the environment are also subject to variability in agreement with figure \ref{fig:cross_corr}. 
In \fbb{} the decoupling scale between dark matter and gas fields shifts from $\sim$1 cMpc$/h$ at $z=6$ to $\sim$5 cMpc$/h$ at $z=0$. We further test whether the model reproduces this behavior in the next section by analyzing the cross-correlation coefficient.
Finally, we emphasize that in practical applications, \embertwo{} can be used to sample many gas realizations for a given dark matter field, in order to asses their respective probability.

\subsection{Statistical and spectral metrics}
In the following we present a number of statistical and spectral metrics to evaluate the networks performance on different testing sets.
On one hand we test the network on the remaining slices of the full hydro run. On the other hand, we also evaluate the network on the corresponding dark-matter only simulation of \fbb{}. This comparison is important since the two dark-matter versions, i.e. from the hydro run (denoted dmh) and from the dark-matter only run (denoted dmo), are expected to differ due to interactions with baryons.
\\
\newline
\textbf{Mass conservation, HI fraction and mass distribution:}
To test the model for mass conservation, HI fraction ($q_\mathrm{HI}$) reconstruction and the probability density function (pdf) we use the slices from the 128 boxes from the test dataset. The 128 variations correspond to the different noise realizations, in order to probe the median and the intrinsic scatter of the learned distribution.
\begin{figure}
    \includegraphics[width=\columnwidth]{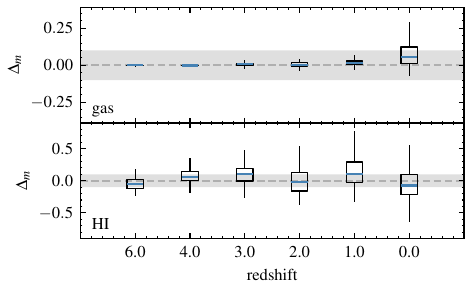}
    \caption{
    Boxplots of the fractional error on the gas mass conservation (top) and HI mass (bottom) across redshifts, where the medians of the distributions are shown in blue. For this test the 128 boxes from the test set were used.
    For each redshift the box and black error bars contain 50 and 90 per cent of the data samples. The grey shaded areas indicate the 10 per cent error bands.
    }
    \label{fig:base_mass}
\end{figure}
From the simulated and emulated denormalized total gas and HI surface densities, we compute the median fractional percentage error on the mass conservation and HI fraction as follows:
\begin{equation}\label{eq:delta_m}
    \Delta_m = \mathrm{median}\left(\frac{m_{p}}{m_{y}} - 1\right), \; \text{where } m_{\phi} = \sum_{ij} \phi_{ij} \cdot A
\end{equation}
and
\begin{equation}\label{eq:delta_qHI}
    \Delta_{q_\mathrm{HI}} = \mathrm{median}\left(\frac{q_{\mathrm{HI},p}}{q_{\mathrm{HI},y}} - 1\right), \; \text{where } q_{\mathrm{HI},\phi}=\frac{\Sigma_{\mathrm{HI}, \phi}}{\Sigma_g, \phi}
\end{equation}
and $A$ denotes the physical area per pixel. $\phi$ refers either to the predicted ($p$) or simulated ($y$) quantity.
In the top panel of figure \ref{fig:base_mass} we show the gas mass conservation at individual redshifts as a boxplot and the 10 per cent level deviation as a grey error band. Overall the reconstructed match closely follows the ground truth of the simulation. For lower redshifts the median (shown in blue) and the scatter in the distributions increases from 0.1 per cent at $z=6$ to $\sim$2.5 per cent at $z=0$ for the total gas mass.
Mass conservation is not guaranteed by construction, neither in individual patches nor in the entire box. This is because the model is trained on individual patches of the simulation and has no information on the entire amount of mass in the box, since this global constraint can not be incorporated into the training process. 
The analysis of the true and predicted pixel pdf of $\Sigma_g$ shown in figure \ref{fig:base_pdf} yields similar results. The largest deviations occur in the high density pixel regime where the mapping becomes more stochastic compared to the lower density regimes. The success of the model is strongly dependent on the normalization scheme, especially for fitting the high density pixel regime, which is crucial for the target of emulating galactic environments.
Combining the findings from figure \ref{fig:base_mass} and \ref{fig:base_pdf} shows that the 2.5 per cent error at $z=0$ stems primarily from the high density regions associated to centers of gaseous halos, where variations of $\sim$2.5 per cent are well within the intrinsic gas mass scatter of $\sim$0.2 dex as shown in \cite{Bernardini2022}.
In figure \ref{fig:base_pdf} we also show the emulated gas pdf when predicted from the dmo input.
This test directly probes how sensitive the emulation results are with respect to the exact nature of the dark matter input. 
The two pdfs are almost identical, showcasing that even though \embertwo{} was originally trained with dmh slices, the dynamic range of gas pixels is well reconstructed from the dmo counterpart.
\\

\begin{figure}
    \includegraphics[width=\columnwidth]{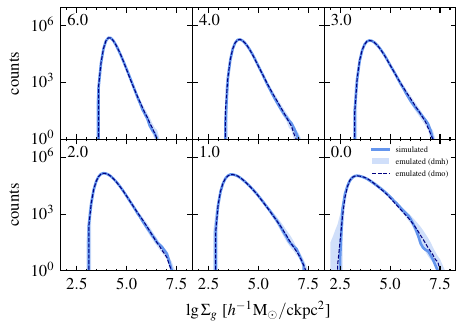}
    \caption{
    Pixel pdf for the simulated and emulated slices of surface gas densities across the entire redshift range (indicated in the top left corners).
    The solid and dashed curves indicate whether the dmh or dmo input was used in the emulation.
    The thick line represents the median of the simulated slices, whereas for the dmh case, the shaded area indicates the 16th to 84th percentiles of the emulated slices of the \fbb{} volume.
    }
    \label{fig:base_pdf}
\end{figure}
Although the model predicts the four baryon channels directly from the dark matter inputs, emulating the HI surface density is much harder compared to the total gas density.
We emphasize that in order to avoid the unphysical case of $q_\mathrm{HI} > 1$, pixel values with exceeding HI fraction are clipped to 1.\footnote{Out of the millions of pixels in our test dataset, this case only occurred for $\sim$10 of them.}
One can observe that the spread of error distribution across redshifts is larger, where the median deviation is typically within 10 per cent and the median error falls inside the distribution for all redshifts in figure \ref{fig:base_mass}.
The spread of the total gas distribution shows a dependence on redshift, indicating that higher redshifts are easier to emulate. For HI on the other hand, there is no clear trend neither in the spread nor the offset of the median. This highlights the complexity of the mapping between dark matter and HI, across the emulated redshift interval of $6 > z > 0$ and is likely attributed to the stronger spatial concentration of HI at lower redshifts. 
In fact, as shown in \cite{Feldmann2023}, below $z=2$ almost the entire HI mass is contained within haloes. In contrast, the total gas field extends more smoothly to larger scales and is thus, conclusively, easier to emulate.\\

\begin{figure}
    \includegraphics[width=\columnwidth]{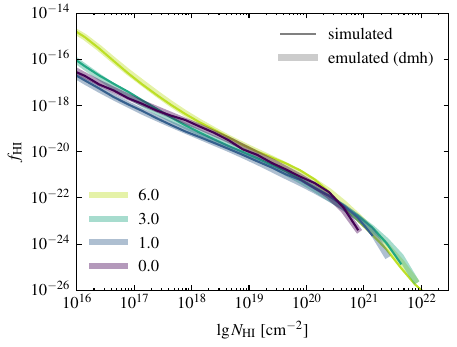}
    \caption{
    Simulated and emulated CDDFs as a function of HI number density for selected redshifts.
    Colors indicate different redshifts given in the bottom left corner. The simulated CDDFs are computed for the entire \fbb{} volume at the specific redshift, while for the emulation we show the median CDDF computed across the 128 emulated test volumes. We do not show the scatter for the emulated boxes, since the curves coincide with the median relation.
    }
    \label{fig:cddf}
\end{figure}

For HI we compute the column density distribution function (CDDF) $f_\mathrm{HI}$ both for the simulated and emulated slices.
The CDDF is a pixel based quantity used in observational HI studies. It is defined such that $f_\mathrm{HI}\diff N_\mathrm{HI} \diff X$ is the number of absorbers per unit column density bin and unit absorption length $\diff X$. Following \cite{Rahmati2013} we write
\begin{equation}
    f_\mathrm{HI} \equiv \frac{\diff^2 n_\mathrm{HI}}{\diff N_\mathrm{HI} \diff X}
\end{equation}
where $n_\mathrm{HI}$ is the number density of HI and $\diff X(z)$ is the absorption distance related to slab thickness $\diff L$ as $\diff X =(H_0/c)(1+z)^{2}\diff L$ (see appendix A in \cite{Bernardini2022} for details).
In figure \ref{fig:cddf} we show the comparison between the simulated and emulated CDDFs for selected redshifts.
The comparison shows that the simulated CDDF is well reconstructed by the \embertwo{} model both in terms of normalization and slope across the redshift range $6\geq z \geq 0$.
Especially, in the lower density regime ($\lg N_\mathrm{HI} \leq 10^{20} \rm cm^{-2}$) the simulated and emulated CDDF agree well with each other. \embertwo{} is capable of reconstructing $f_\mathrm{HI}$ of these column densities with almost no information loss. 
For column densities beyond $\lg N_\mathrm{HI} \geq 10^{20} \rm cm^{-2}$ the CDDF steepens with an exponential cutoff at higher density systems. \cite{Rahmati2013} showed that in order to have a converged CDDF up to $10^{22}\rm cm^{-2}$ at $z=0$ one requires a box size of at least 50 Mpc/h.
Although the CDDFs might not be fully converged in this high density regime, the simulated and emulated slopes of the CDDF still agree almost perfectly with each other.
\\

\begin{figure}
    \includegraphics[width=\columnwidth]{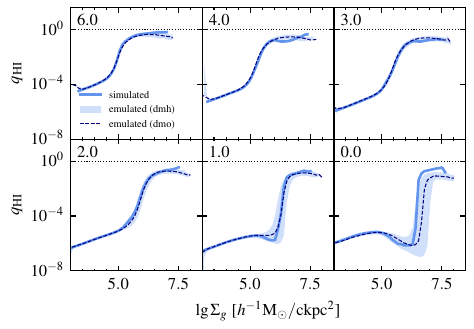}
    \caption{
    Comparison of the HI fraction $q_\mathrm{HI}$ as a function of gas surface density for different redshifts (indicated in the top left corners).
    The solid and dashed curves indicate whether the dmh or dmo input was used in the emulation.
    The thick line represents the median of the simulated slices, whereas for the dmh case, the shaded area indicates the 16th to 84th percentiles of the emulated slices of the \fbb{} volume.
    }
    \label{fig:base_qHI}
\end{figure}

The mass conservation analysis of total gas and HI as well as the CDDF only represent global summary statistics. They do not answer the question of how well the distribution of HI within the gas reservoirs is reconstructed.
To this end, we compare the simulated and emulated HI fractions as a function of the gas surface density in figure \ref{fig:base_qHI}. However, we emphasize that in contrast to the mass error analysis, we compute the HI fraction of each pixel and aggregate the statistic by taking the median of that distribution as opposed to computing the HI fractions on the box level.
For the simulation we show the median relation, whereas the 16th to 84th percentile is shown for the emulated slices. 
For most regimes and redshifts, \embertwo{} emulates realistic HI fractions that fall within the emulated distribution.
The overall evolution of the reconstructed HI fractions highlights the fact that for $z\leq 1$ is highly concentrated in high gas density peaks.
The largest discrepancy occurs at $z=0$ where the simulated $q_\mathrm{HI}$ bends upwards at $\Sigma_g\simeq 10^6$ $h^{-1}$M$_\odot/$ckpc$^2$ whereas for the emulation this feature is systematically shifted to higher densities by $\sim$0.5 dex.
Similarly, at $z=0$, $q_\mathrm{HI}$ is systematically underpredicted by $\sim$0.5 dex for $\Sigma_g\simeq 10^7$ $h^{-1}$M$_\odot/$ckpc$^2$.
However, we emphasize that by taking into consideration the gas pdf shown in figure \ref{fig:base_pdf}, this behavior of the model only affects an order of $\sim$100 pixels.
These pixels represent halo centers with large reservoirs of HI, which are especially interesting in the context of galaxy formation and their accuracy should be weighted more in the emulation process.
Future refinements of the model will be geared towards reproducing the HI fractions in these extreme cases with higher accuracy.
Similarly to the results presented in figure \ref{fig:base_pdf}, the networks predictions are nearly insensitive to whether the dmh or dmo inputs were used in the emulation, since the median relation for the dmo emulation falls inside the percentiles region of the dmh emulation for all redshifts. Hence, the robustness of \embertwo{} with respect to dmo or dmh inputs is valid even for the emulated HI channel.
\\

\begin{figure}
    \includegraphics[width=\columnwidth]{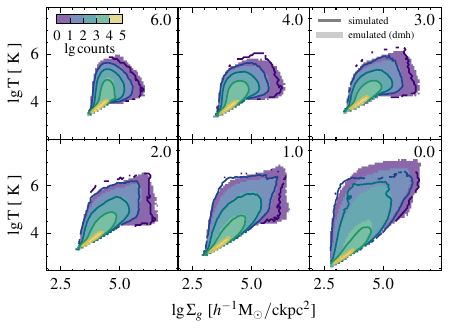}
    \caption{
    Comparison of the density-temperature diagram for total gas surface density in \fbb{} (solid contours) and emulated (shaded contours) counterparts by \embertwo{}. The different redshifts are indicated in the top right corner of each panel. For the simulation the figure represents the median histogram of all slices in the \fbb{} volume. 
    For the emulation we show the median of all slices of the 128 test boxes.
    }
    \label{fig:phase_gas}
\end{figure}

\noindent \textbf{Density-temperature diagrams:} 
The mass conservation and HI fraction analysis revealed that the model overall reproduces realistic HI distributions. 
In addition, we can add the temperature information to test how well different gas-temperature phases are emulated across the redshift range.
Figure \ref{fig:phase_gas} shows the phase diagrams of simulated and emulated total gas surface density as a function of temperature where the colors indicate the discrete height of the histograms for better comparison.

Generally, the model predictions are in very good agreement with the simulation across the entire redshift range. An example of that is the excellent reproduction of the sharp cut-offs at lower temperatures of $T\lesssim 10^4$ K, i.e. there is only little contamination in the emulated fields and the lower thresholds is sharply reproduced across redshifts and temperatures for all gas densities ranging from the diffuse IGM to the denser and colder halo gas ($T \lesssim 10^4$ K and $\Sigma_\mathrm{gas} \gtrsim 10^6 h^{-1} \rm M_\odot / ckpc^2$).
We emphasize that due to projection effects in the maps, hot outflows can overlap with halo centers, i.e. some pixels show large total gas surface densities with high temperatures.

A minor discrepancy between simulation and emulation is the region of the Hot Medium (HM) with $T > 10^7$ K and the cooler but denser Warm CGM (WCGM) with densities $\rm \Sigma_g \gtrsim 10^6$ $h^{-1}$ $\rm M_\odot/ckpc^2$. 
In these parts of the phase-space diagram the model predicts that parts of the distribution reach high temperatures $\gtrsim 10^7$ K.
We have investigated the origin of these gas cells, and found that they correspond to gas cells in extreme outflows in and around gaseous haloes where gas is compressed and heated. In rare occasions the model overestimates the abundance of these dense, high temperature gas pixels in the HM. Furthermore, the contamination in this regime is primarily a temperature related issue, i.e. the predicted density is correct but the temperature is too high resulting in cells that are shifted to the top in the distribution.
We emphasize however that on the field level this is a small effect, since the fraction of these pixels is less than 1 per cent of the entire distribution.\\

\noindent \textbf{Cross-correlations:} A robust spectral metric for the presented machine learning application is a fundamental tool to measure the capabilities of the neural network.
Similar works \citep[e.g.][]{Troester2019, Wadekar2021} that train emulators on the field level adopt the spatial density power spectrum as a fundamental metric to analyse higher order moments of predicted and true fields.
Due to the small boxsize of the simulations in our training set, it turns out that e.g. the power spectrum is not a good metric, since for small cosmological volumes it is disproportionally sensitive to individual high density pixels in the simulation. 
\cite{vanDaalen2019} found that massive haloes significantly contribute to the power spectrum on scales $k \geq 10 \, h / \text{cMpc}$. This was also observed by \cite{Chisari2018} by comparing power spectra from sub-volumes drawn from a $100 \, \text{cMpc}/h$ simulation. The same effect is also present in smaller cosmological volumes of $25 \, \text{cMpc}/h$ boxes \citep[]{Nicola2022, Delgado2023, Gebhardt2024}{}.
Depending on whether a massive object is present in a given sub-volume or not, the power spectrum varies significantly between samples.
However, it is a priori not clear whether other quantities like the spatial cross-power spectrum (here between dark matter and total gas and HI densities) and the cross-correlation coefficient are equally affected.
To clarify this aspect, we design the following test to quantify how suitable the power spectrum $P$, the cross-power spectrum $P^{\times}$ and the cross-correlation coefficient $r$ are as spectral metrics for our approach.

\begin{figure}
    \includegraphics[width=\columnwidth]{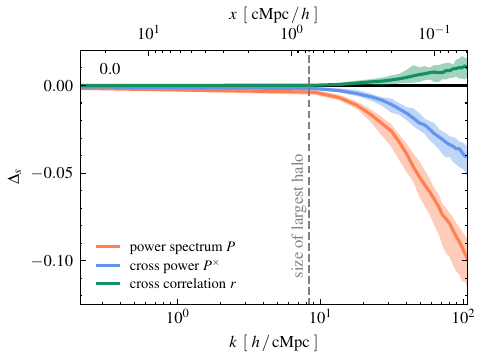}
    \caption{
    Relative percentage error, $\Delta_s$, of the spatial power spectrum $P$, cross-power spectrum $P^{\times}$ and cross-correlation coefficient $r$ for the simulated and modified slices of the \fbb{} simulation at $z=0$. For the modified slices, the gas mass of the five most massive pixels was reduced by 10 per cent. Coloured lines and shaded regions denote the median and the 16th to 84th percentile across all slices.
    The cross-correlation coefficient is almost unchanged by this modification, while the cross-power and especially the power spectrum change by orders of 4 per cent and 10 per cent rendering them unusable as a metric for our spectral analysis of the network. We emphasize that only simulated data has been used for this test.
    }
    \label{fig:power_study}
\end{figure}

We take the 2-dimensional slices from the \fbb{} simulation and identify the 5 (out of 1024$^2$) largest gas mass pixels (hereafter denoted as 'modified' pixels).
The total gas mass of the modified pixels is then manually changed via
\begin{equation}
    m_g \rightarrow m_g \times 0.9,
\end{equation}
i.e. we remove 10 per cent of the pixels gas mass. Since the hot pixels are representing galaxies located at the centers of haloes, 10 per cent deviation in mass lies within the intrinsic scatter of the halo-mass to gas-mass relation, see e.g. \cite{Bernardini2022} for the \fb{} simulation or \cite{Stiskalek2022} for the IllustrisTNG simulation. This is why we chose a value of 10 per cent for this test. 
We then compute $P$, $P^{\times}$ and $r$ for the simulated as well as for the modified slices and measure the fractional percentage error 
\begin{equation}
    \Delta_s = \mathrm{median}\left( \frac{s_\mathrm{mod}}{s_\mathrm{sim}} - 1 \right),
\end{equation}
where $s=P, P^{\times}, r$.
The result is shown for $z=0$ in figure \ref{fig:power_study}. The power and cross-power spectra are heavily affected with deviations reaching up to 10 per cent and 4 per cent respectively. Moreover, modifying the maps is not just affecting the smallest pixel scale, but changes the spectra down to scales of several hundred ckpc$/h$ reaching the physical scale of the largest halo in the \fbb{} simulation.
Power and cross-power spectra might be suitable metrics for larger simulation boxes or other field-level applications, but do not represent meaningful statistics for our approach.
However, the cross-correlation coefficient $r$ appears to be robust against these modifications, that mimic realistic changes in the gas mass (maximum error is $\sim$1 per cent). Thus, we decide to use it as our main metric when evaluating the spectral capabilities of the network.\\

\begin{figure}
    \includegraphics[width=\columnwidth]{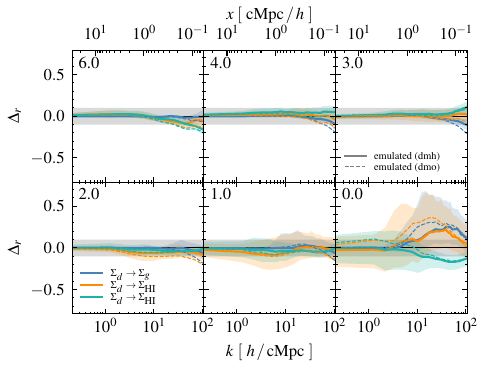}
    \caption{
    Cross-correlation errors $\Delta_r$ across scales and redshifts (indicated in the upper left corners) between the emulated fields from \embertwo{} and the simulated \fbb{} fields. Shown are three cross-correlations between the input dark matter field and the output gas and HI surface densities, as well as the cross-correlation between the gas and HI surface density fields in green.
    The solid and dashed curves indicate whether the dmh or dmo input was used in the emulation.
    For the dmh case, the shaded area and solid lines represent the median and 16th to 84th percentile at each $k$-scale, while the grey band indicates the 10 per cent error.
    For better visibility, we do not show additional percentiles for the dmo case.}
    \label{fig:cross_corr_dm}
\end{figure}

\noindent\textbf{Cross-correlations of emulated fields:} We use the same 128 test boxes at each redshift, to compare simulated and emulated cross-correlation coefficients. 
Similar to the percentage errors for mass and HI fraction, we define the error on $r(k)$ as
\begin{equation}
    \Delta_r = \mathrm{median}\left(
    \frac{r_{p}}{r_{y}} - 1 
    \right),
\end{equation}
where $r=r(k)$ is defined in equation \ref{eq:r}.
Figure \ref{fig:cross_corr_dm} displays the error $\Delta_r$ between the simulated and emulated cross-correlation coefficients computed by correlating the dark matter and the total gas surface density (blue).
We also show the cross-correlation error between dark matter and HI (orange).
We emphasize that the intrinsic scatter of the presented cross-correlations in the simulation is $\sim$10 per cent, which we use to quantify the networks performance.
For all redshifts except $z=0$ these cross-correlation errors and their median fall within the 10 per cent uncertainty region, indicating that the model reproduces the two-point statistics between input and output across all redshifts.
The behavior is the same whether the dmo or dmh input is used in the emulation.
For $z=0$ the statistics deviate a maximum of $\sim$20 per cent in line with the findings in previous paragraphs.
Additionally, we also show the cross-correlation error between model and simulation between the two gas density fields, i.e. total gas and HI (shown in green). The deviation in this case is within $\sim$20 per cent. This behavior shows that although \embertwo{} overpredicts $r$ both for total gas and HI, the two emulated baryon fields are still statistically sound in terms of cross-correlations even at $z=0$, independent of whether the dmo or dmh inputs are used.
For lower redshifts ($z<1$) we hypothesize that an even larger training and testing set might improve the errors, since for these lower redshifts the boxsize and thus the amount of cosmic structure becomes an important factor that impacts the cross-correlation between dark matter and gas.
Additionally, figure \ref{fig:cross_corr_dm} showcases that the intrinsic scatter in the predicted cross-correlations increases with redshift. 
The reason for this behavior is two-fold. On one hand, this increase is expected, since the scatter of the cross-correlation also increases at lower $z$ as shown in figure \ref{fig:cross_corr}, i.e. the scatter is intrinsic to the dataset. 
On the other hand, some of the uncertainty is inherent to the model, since less data samples are available at lower redshifts, and thus in extreme cases the model is less confident.\\

\begin{figure}
    \includegraphics[width=\columnwidth]{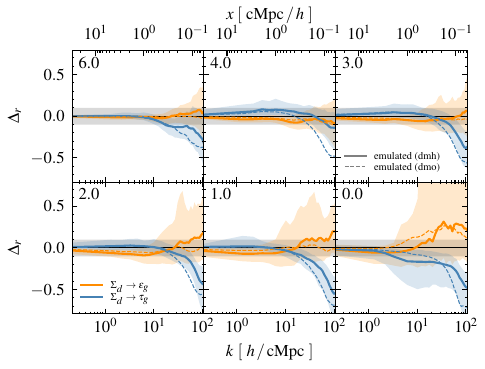}
    \caption{
    Cross-correlation errors $\Delta_r$ across scales and redshifts (indicated in the upper left corners) between the emulated fields from \embertwo{} and the simulated \fbb{} fields. Shown are the cross-correlation errors for the kinetic and thermal surface energy densities $\varepsilon_g$ and $\tau_g$.
    The solid and dashed curves indicate whether the dmh or dmo input was used in the emulation.
    The shaded area and solid lines represent the median and 16th to 84th percentile at each $k$-scale, while the grey band indicates the 10 per cent error.
    For better visibility, we do not show additional percentiles for the dmo case.}
    \label{fig:cross_corr_derived_fields}
\end{figure}

\noindent\textbf{Cross-correlations of derived fields:} The analysis presented in figure \ref{fig:cross_corr_dm} shows that \embertwo{} has not only learned the stochastic mapping between dark matter and gas densities, but the two channels of gas densities are statistically compatible.
Similarly, we show in figure \ref{fig:cross_corr_derived_fields} that the additional emulated channels of baryons are also reproduced reasonably well and in a statistically sound manner. 
We define two energy fields, derived from the emulation outputs, namely
\begin{align}
    \varepsilon_g &= \Sigma_g \times v^2_g/2 \\
    \tau_g &= \Sigma_g \times k_\mathrm{B} T
\end{align}
where $\varepsilon_g$ can be understood as a kinetic energy surface density, $\tau_g$ is proportional to the thermal energy density and $k_\mathrm{B}$ is the Boltzmann constant. 
Figure \ref{fig:cross_corr_derived_fields} shows the cross-correlation error $\Delta_r$ between the derived fields when correlating them with the dark matter input density. This analysis is of particular interest, since we combine two emulated fields together and correlate them with the dark matter density input. If the model does not capture the cross-correlation between the different fields, and especially across baryon channels, this behavior would be visible in the cross-correlation analysis.
The kinetic energy density shown in orange is reproduced well across all scales and redshifts with maximum deviations of order 10 per cent but lies well within the 32 to 68 percentiles. 
$\tau_g$, which includes the temperature field, shows more severe systematic deviations. We have argued that emulating the temperature field at lower redshifts is harder due to strong outflows in the simulation, which is why outliers in the spectrum are expected. 
Throughout the redshift range, the cross-correlation of the thermal energy density deviates above a certain decoupling scale, which shifts to smaller $k$ values for decreasing redshift.
At $z=6$ the deviations at the smallest scale of $\sim$100 ckpc$/h$ is of order 10 per cent. This scale shifts to $\sim$1 cMpc$/h$ at $z=0$. The systematic behavior of underestimating the power above the decoupling scales indicates that the neural network is not able to capture the exact correlations between $\Sigma_g$ and $T$, such that combining these two fields yields predictions that accurately represent the physics in \fbb{} only below the decoupling scales. 
This drop in cross-power is present in both dmo and dmh emulations, although while the dmo predictions show the same scale of decoupling, they then typically fall off faster compared to the dmh predictions.
Typically, $T$ is lower in void and intermediate density regions, correlating positively with $\Sigma_g$, i.e. $\Sigma_g$ and $T$ increase jointly. 
However, while $\Sigma_g$ has sharp density contrasts around filaments and haloes, the high temperature field typically extends around them, rendering the correlation between density and temperature more complex on scales between $\sim$1 cMpc$/h$ and $\sim$100 ckpc$/h$.
Finally, on scales below $\sim$100 ckpc$/h$, shock-heated gas cools inside haloes and the temperature and density profiles anti-correlate increasingly towards the halo centers.
These arguments highlight that the temperature field of the CGM is the most challenging field to model for \embertwo{} where additional improvements are needed.


\section{Conclusions}\label{sec:conclusion}
In this paper we presented \embertwo{}, an improved neural network architecture to emulate baryons from dark matter on the field level over the continuous redshift range of $z \leq 6 \leq 0$.
\textsf{EMBER-2} is designed to predict 2-dimensional gas fields including density, radial velocity, temperature and HI densities from dark matter density and radial velocity at a pixel resolution of $\sim$29 ckpc$/h$ and a depth resolution of 1.5 cMpc$/h$. 
In the following we will summarize the \embertwo{} model and its predictions of the emulated gas properties.
\begin{itemize}
    \item The model has successfully learned to interpolate the data distribution across a large range of redshifts, from $z=6$ to $z=0$. 
    By probing the distribution we found that the convolution kernels of the network are modulated according to the redshift variable, allowing to emulate the dark matter patches $x$ smoothly across cosmic times with no memory bottleneck. (first two rows in figure \ref{fig:base_visuals})
    
    \item In a similar analysis we interpolated the noise parameter $\eta$ to understand the range of scales that are affected by the noise injection scheme.
    We found that the noise variable is indeed a crucial ingredient in the emulation of cosmic structure, and that the model has learned the importance of incorporating noise across the entire redshift range. At early times, $\eta$ impacts mostly individual high density peaks in the field, whereas at lower redshifts noise is used to emulate features of cosmic structure across all scales from tens of kpc to Mpc. (third row in figure \ref{fig:base_visuals})
    
    \item The model is capable of emulating new realizations of gas slices that are consistent regarding the global emulated gas mass and HI fraction. 
    This result is independent of whether the dmo or dmh inputs are used.
    The median errors are within 5 per cent for the total gas mass and 10 per cent for the HI mass across the entire redshift range. (figure \ref{fig:base_mass})
    
    \item We quantified the spectral agreement of the model with the simulation by means of the cross-correlation coefficient between gas and dark matter. 
    The error of the model is within the 5 per cent margin throughout the entire dataset except for the $z=0$ snapshot where the errors are of order 20 per cent. This result is independent of whether the dmo or dmh inputs are used. (figure \ref{fig:cross_corr_dm})

    \item Moreover, the \embertwo{} model emulates total gas densities, radial velocities and HI density fields in a statistically sound manner. To quantify the spectral coherence between baryon channels, we showed that the cross-correlation errors on the derived kinetic energy density is within 10 per cent uncertainty both for the dmo and dmh emulations. 
    For the thermal energy density, \embertwo{} matches well with the simulation counterparts, but systematically underestimates the power above a decoupling scale.
\end{itemize}

\noindent \textbf{Future improvements:} As presented in this work, the \embertwo{} framework is designed to learn mappings on the field level between two domains; here between dark matter and baryons.
In future work we will investigate to what extent additional input fields such as the dark matter velocity dispersion or gradients of dark matter densities would improve the predictive power of our algorithm, as e.g. dynamical fields can encode signals of recent merger events \citep{Hoeft2004}.
Including more fields is also possible on the baryon side, where we will investigate the capabilities of \embertwo{} to emulate more gas properties, such as e.g. metallicities.
Further, we will investigate how to improve the accuracies for the HI fraction at $z=0$ as well as how to emulate more statistically sound temperature fields. 
To this end, additional changes to the architecture are needed to capture the small scale correlations more accurately.
One possible approach is to strengthen the individual ModConv blocks through architecture changes, such as e.g. the \textit{Attention} mechanism first introduced by \cite{Vasvani_2017}. This particular approach however has major memory drawbacks \citep[e.g.][]{Zhang2_2018}, especially for emulating large cosmic maps, hence more investigation in this direction is needed in the future.

The central methodological change of \embertwo{} is the use of the ModConv scheme, which allows to parameterize the convolution kernels based on embeddings that are learned from global variables, such as the redshift.
Presumably, the same factorization framework will allow the network to learn smooth interpolations over any input parameter space given a large enough and densely spaced training set.
Future work will investigate the interpolation capabilities of the model when presented with simulations with varying astrophysical parameters, e.g. different stellar or AGN feedback strengths encoded through the global context variables.
Due to the small inference times, such an emulator, based on the \embertwo{} model, could directly be paired with sampling algorithms (e.g. MCMC) and observations to perform parameter inference to constrain the physics of galaxy formation.


\section*{Acknowledgments}
The \fb{} simulation was supported in part by computing allocations at the Swiss National Supercomputing Centre (CSCS) under project IDs s697, s698, and uzh18. 
The \fbb{}{} simulation was supported by computing allocations at CSCS under project IDs s1255 and uzh18.
MB and RF acknowledge financial support from the Swiss National Science Foundation (grant no 200021\_188552 and CRSII5\_193826).
JG gratefully acknowledges financial support from the Swiss National Science Foundation (grant no CRSII5\_193826).
DAA acknowledges support by NSF grant AST-2108944, NASA grant ATP23-0156, STScI grants JWST-GO-01712.009-A and JWST-AR-04357.001-A, Simons Foundation Award CCA-1018464, and Cottrell Scholar Award CS-CSA-2023-028 by the Research Corporation for Science Advancement.
CAFG was supported by NSF through grants AST-2108230 and AST-2307327; by NASA through grant 21-ATP21-0036; and by STScI through grant JWST-AR-03252.001-A.
RB is supported by the SNSF through the Ambizione Grant PZ00P2-223532.
We thank the anonymous referee for his valuable comments that helped improve this work.
This work made use of infrastructure services provided by \href{www.s3it.uzh.ch}{S3IT}, the Service and Support for Science IT team at the University of Zurich.

\section*{Data Availability Statement}
We provide the source code at the official repository: \url{https://github.com/maurbe/ember2}.
The data used to produce the plots will be made available upon reasonable request to the corresponding author.


\bibliographystyle{mnras}
\bibliography{document}

\begin{thebibliography}{}
\makeatletter
\relax
\def\mn@urlcharsother{\let\do\@makeother \do\$\do\&\do\#\do\^\do\_\do\%\do\~}
\def\mn@doi{\begingroup\mn@urlcharsother \@ifnextchar [ {\mn@doi@} {\mn@doi@[]}}
\def\mn@doi@[#1]#2{\def\@tempa{#1}\ifx\@tempa\@empty \href {http://dx.doi.org/#2} {doi:#2}\else \href {http://dx.doi.org/#2} {#1}\fi \endgroup}
\def\mn@eprint#1#2{\mn@eprint@#1:#2::\@nil}
\def\mn@eprint@arXiv#1{\href {http://arxiv.org/abs/#1} {{\tt arXiv:#1}}}
\def\mn@eprint@dblp#1{\href {http://dblp.uni-trier.de/rec/bibtex/#1.xml} {dblp:#1}}
\def\mn@eprint@#1:#2:#3:#4\@nil{\def\@tempa {#1}\def\@tempb {#2}\def\@tempc {#3}\ifx \@tempc \@empty \let \@tempc \@tempb \let \@tempb \@tempa \fi \ifx \@tempb \@empty \def\@tempb {arXiv}\fi \@ifundefined {mn@eprint@\@tempb}{\@tempb:\@tempc}{\expandafter \expandafter \csname mn@eprint@\@tempb\endcsname \expandafter{\@tempc}}}

\bibitem[\protect\citeauthoryear{Ade et~al.,}{Ade et~al.}{2016}]{Planck2016}
Ade P. A.~R.,  et~al., 2016, \mn@doi [Astronomy & Astrophysics] {10.1051/0004-6361/201525830}, 594, A13

\bibitem[\protect\citeauthoryear{Adler \& Öktem}{Adler \& Öktem}{2018}]{Adler2018}
Adler J.,  Öktem O.,  2018 (\mn@eprint {arXiv} {1811.05910})

\bibitem[\protect\citeauthoryear{{Angl{\'e}s-Alc{\'a}zar}, {Dav{\'e}}, {{\"O}zel}  \& {Oppenheimer}}{{Angl{\'e}s-Alc{\'a}zar} et~al.}{2014}]{Angles2014}
{Angl{\'e}s-Alc{\'a}zar} D.,  {Dav{\'e}} R.,  {{\"O}zel} F.,   {Oppenheimer} B.~D.,  2014, \mn@doi [\apj] {10.1088/0004-637X/782/2/84}, \href {https://ui.adsabs.harvard.edu/abs/2014ApJ...782...84A} {782, 84}

\bibitem[\protect\citeauthoryear{{Angl{\'e}s-Alc{\'a}zar}, {Faucher-Gigu{\`e}re}, {Kere{\v{s}}}, {Hopkins}, {Quataert}  \& {Murray}}{{Angl{\'e}s-Alc{\'a}zar} et~al.}{2017}]{Angles2017}
{Angl{\'e}s-Alc{\'a}zar} D.,  {Faucher-Gigu{\`e}re} C.-A.,  {Kere{\v{s}}} D.,  {Hopkins} P.~F.,  {Quataert} E.,   {Murray} N.,  2017, \mn@doi [\mnras] {10.1093/mnras/stx1517}, \href {https://ui.adsabs.harvard.edu/abs/2017MNRAS.470.4698A} {470, 4698}

\bibitem[\protect\citeauthoryear{{Aric{\`o}}, {Angulo}, {Contreras}, {Ondaro-Mallea}, {Pellejero-Iba{\~n}ez}  \& {Zennaro}}{{Aric{\`o}} et~al.}{2021}]{Arico2021}
{Aric{\`o}} G.,  {Angulo} R.~E.,  {Contreras} S.,  {Ondaro-Mallea} L.,  {Pellejero-Iba{\~n}ez} M.,   {Zennaro} M.,  2021, \mn@doi [\mnras] {10.1093/mnras/stab1911}, \href {https://ui.adsabs.harvard.edu/abs/2021MNRAS.506.4070A} {506, 4070}

\bibitem[\protect\citeauthoryear{Arjovsky \& Bottou}{Arjovsky \& Bottou}{2017}]{Arjovsky2017a}
Arjovsky M.,  Bottou L.,  2017, Towards Principled Methods for Training Generative Adversarial Networks (\mn@eprint {arXiv} {1701.04862})

\bibitem[\protect\citeauthoryear{Arjovsky, Chintala  \& Bottou}{Arjovsky et~al.}{2017}]{Arjovsky2017b}
Arjovsky M.,  Chintala S.,   Bottou L.,  2017, Wasserstein GAN (\mn@eprint {arXiv} {1701.07875})

\bibitem[\protect\citeauthoryear{{Balsara}, {Livio}  \& {O'Dea}}{{Balsara} et~al.}{1994}]{Balsara1994}
{Balsara} D.,  {Livio} M.,   {O'Dea} C.~P.,  1994, \mn@doi [\apj] {10.1086/174977}, \href {https://ui.adsabs.harvard.edu/abs/1994ApJ...437...83B} {437, 83}

\bibitem[\protect\citeauthoryear{Barber, Starkenburg, Navarro  \& McConnachie}{Barber et~al.}{2014}]{Barber2014}
Barber C.,  Starkenburg E.,  Navarro J.~F.,   McConnachie A.~W.,  2014, \mn@doi [\mnras] {10.1093/mnras/stu2494}, 447, 1112

\bibitem[\protect\citeauthoryear{{Bassini}, {Feldmann}, {Gensior}, {Faucher-Gigu{\`e}re}, {Cenci}, {Moreno}, {Bernardini}  \& {Liang}}{{Bassini} et~al.}{2024}]{Bassini2024}
{Bassini} L.,  {Feldmann} R.,  {Gensior} J.,  {Faucher-Gigu{\`e}re} C.-A.,  {Cenci} E.,  {Moreno} J.,  {Bernardini} M.,   {Liang} L.,  2024, \mn@doi [\mnras] {10.1093/mnrasl/slae036}, \href {https://ui.adsabs.harvard.edu/abs/2024MNRAS.tmpL..35B} {}

\bibitem[\protect\citeauthoryear{{Bernardini}, {Feldmann}, {Angl{\'e}s-Alc{\'a}zar}, {Boylan-Kolchin}, {Bullock}, {Mayer}  \& {Stadel}}{{Bernardini} et~al.}{2022}]{Bernardini2022}
{Bernardini} M.,  {Feldmann} R.,  {Angl{\'e}s-Alc{\'a}zar} D.,  {Boylan-Kolchin} M.,  {Bullock} J.,  {Mayer} L.,   {Stadel} J.,  2022, \mn@doi [\mnras] {10.1093/mnras/stab3088}, \href {https://ui.adsabs.harvard.edu/abs/2022MNRAS.509.1323B} {509, 1323}

\bibitem[\protect\citeauthoryear{{Bieri}, {Naab}, {Geen}, {Coles}, {Pakmor}  \& {Walch}}{{Bieri} et~al.}{2023}]{Bieri2023}
{Bieri} R.,  {Naab} T.,  {Geen} S.,  {Coles} J.~P.,  {Pakmor} R.,   {Walch} S.,  2023, \mn@doi [\mnras] {10.1093/mnras/stad1710}, \href {https://ui.adsabs.harvard.edu/abs/2023MNRAS.523.6336B} {523, 6336}

\bibitem[\protect\citeauthoryear{Biernacki \& Teyssier}{Biernacki \& Teyssier}{2018}]{Biernacki2018}
Biernacki P.,  Teyssier R.,  2018, \mn@doi [\mnras] {10.1093/mnras/sty216}, 475, 5688–5703

\bibitem[\protect\citeauthoryear{{Birnboim} \& {Dekel}}{{Birnboim} \& {Dekel}}{2003}]{Birnboim_2003}
{Birnboim} Y.,  {Dekel} A.,  2003, \mn@doi [\mnras] {10.1046/j.1365-8711.2003.06955.x}, \href {https://ui.adsabs.harvard.edu/abs/2003MNRAS.345..349B} {345, 349}

\bibitem[\protect\citeauthoryear{{Bluck} et~al.,}{{Bluck} et~al.}{2019}]{Bluck2019}
{Bluck} A. F.~L.,  et~al., 2019, \mn@doi [\mnras] {10.1093/mnras/stz363}, \href {https://ui.adsabs.harvard.edu/abs/2019MNRAS.485..666B} {485, 666}

\bibitem[\protect\citeauthoryear{{Borrow}, {Angl{\'e}s-Alc{\'a}zar}  \& {Dav{\'e}}}{{Borrow} et~al.}{2020}]{Borrow2020}
{Borrow} J.,  {Angl{\'e}s-Alc{\'a}zar} D.,   {Dav{\'e}} R.,  2020, \mn@doi [\mnras] {10.1093/mnras/stz3428}, \href {https://ui.adsabs.harvard.edu/abs/2020MNRAS.491.6102B} {491, 6102}

\bibitem[\protect\citeauthoryear{{Bottrell}, {Torrey}, {Simard}  \& {Ellison}}{{Bottrell} et~al.}{2017}]{Bottrell2017}
{Bottrell} C.,  {Torrey} P.,  {Simard} L.,   {Ellison} S.~L.,  2017, \mn@doi [\mnras] {10.1093/mnras/stx276}, \href {https://ui.adsabs.harvard.edu/abs/2017MNRAS.467.2879B} {467, 2879}

\bibitem[\protect\citeauthoryear{{Bournaud}, {Elmegreen}, {Teyssier}, {Block}  \& {Puerari}}{{Bournaud} et~al.}{2010}]{Bournaud2019}
{Bournaud} F.,  {Elmegreen} B.~G.,  {Teyssier} R.,  {Block} D.~L.,   {Puerari} I.,  2010, \mn@doi [\mnras] {10.1111/j.1365-2966.2010.17370.x}, \href {https://ui.adsabs.harvard.edu/abs/2010MNRAS.409.1088B} {409, 1088}

\bibitem[\protect\citeauthoryear{{Brooks}, {Governato}, {Quinn}, {Brook}  \& {Wadsley}}{{Brooks} et~al.}{2009}]{Brooks_2009}
{Brooks} A.~M.,  {Governato} F.,  {Quinn} T.,  {Brook} C.~B.,   {Wadsley} J.,  2009, \mn@doi [\apj] {10.1088/0004-637X/694/1/396}, \href {https://ui.adsabs.harvard.edu/abs/2009ApJ...694..396B} {694, 396}

\bibitem[\protect\citeauthoryear{{Bullock}, {Kravtsov}  \& {Weinberg}}{{Bullock} et~al.}{2001}]{Bullock2001}
{Bullock} J.~S.,  {Kravtsov} A.~V.,   {Weinberg} D.~H.,  2001, \mn@doi [\apj] {10.1086/318681}, \href {https://ui.adsabs.harvard.edu/abs/2001ApJ...548...33B} {548, 33}

\bibitem[\protect\citeauthoryear{{Cenci}, {Feldmann}, {Gensior}, {Moreno}, {Bassini}  \& {Bernardini}}{{Cenci} et~al.}{2024a}]{Cenci2024a}
{Cenci} E.,  {Feldmann} R.,  {Gensior} J.,  {Moreno} J.,  {Bassini} L.,   {Bernardini} M.,  2024a, \mn@doi [\mnras] {10.1093/mnras/stad3709}, \href {https://ui.adsabs.harvard.edu/abs/2024MNRAS.527.7871C} {527, 7871}

\bibitem[\protect\citeauthoryear{{Cenci}, {Feldmann}, {Gensior}, {Bullock}, {Moreno}, {Bassini}  \& {Bernardini}}{{Cenci} et~al.}{2024b}]{Cenci2024b}
{Cenci} E.,  {Feldmann} R.,  {Gensior} J.,  {Bullock} J.~S.,  {Moreno} J.,  {Bassini} L.,   {Bernardini} M.,  2024b, \mn@doi [\apjl] {10.3847/2041-8213/ad1ffb}, \href {https://ui.adsabs.harvard.edu/abs/2024ApJ...961L..40C} {961, L40}

\bibitem[\protect\citeauthoryear{Chandrasegaran, Tran  \& Cheung}{Chandrasegaran et~al.}{2021}]{Chandrasegaran2021}
Chandrasegaran K.,  Tran N.-T.,   Cheung N.-M.,  2021 (\mn@eprint {arXiv} {2103.17195})

\bibitem[\protect\citeauthoryear{Chen, Li, Jin, Liu  \& Li}{Chen et~al.}{2020}]{Chen2020}
Chen Y.,  Li G.,  Jin C.,  Liu S.,   Li T.,  2020 (\mn@eprint {arXiv} {2012.05535})

\bibitem[\protect\citeauthoryear{Chisari et~al.,}{Chisari et~al.}{2018}]{Chisari2018}
Chisari N.~E.,  et~al., 2018, \mn@doi [\mnras] {10.1093/mnras/sty2093}, 480, 3962–3977

\bibitem[\protect\citeauthoryear{{Choi}, {Ostriker}, {Naab}, {Oser}  \& {Moster}}{{Choi} et~al.}{2015}]{Choi2015}
{Choi} E.,  {Ostriker} J.~P.,  {Naab} T.,  {Oser} L.,   {Moster} B.~P.,  2015, \mn@doi [\mnras] {10.1093/mnras/stv575}, \href {https://ui.adsabs.harvard.edu/abs/2015MNRAS.449.4105C} {449, 4105}

\bibitem[\protect\citeauthoryear{{Cohen} \& {Welling}}{{Cohen} \& {Welling}}{2016}]{Cohen_2016}
{Cohen} T.~S.,  {Welling} M.,  2016, Group Equivariant Convolutional Networks (\mn@eprint {arXiv} {1602.07576})

\bibitem[\protect\citeauthoryear{Cohen, Fialkov, Barkana  \& Monsalve}{Cohen et~al.}{2020}]{Cohen2020}
Cohen A.,  Fialkov A.,  Barkana R.,   Monsalve R.~A.,  2020, \mn@doi [\mnras] {10.1093/mnras/staa1530}, 495, 4845–4859

\bibitem[\protect\citeauthoryear{{Correa}, {Schaye}, {Clauwens}, {Bower}, {Crain}, {Schaller}, {Theuns}  \& {Thob}}{{Correa} et~al.}{2017}]{Correa2017}
{Correa} C.~A.,  {Schaye} J.,  {Clauwens} B.,  {Bower} R.~G.,  {Crain} R.~A.,  {Schaller} M.,  {Theuns} T.,   {Thob} A. C.~R.,  2017, \mn@doi [\mnras] {10.1093/mnrasl/slx133}, \href {https://ui.adsabs.harvard.edu/abs/2017MNRAS.472L..45C} {472, L45}

\bibitem[\protect\citeauthoryear{{Correa}, {Schaye}, {Wyithe}, {Duffy}, {Theuns}, {Crain}  \& {Bower}}{{Correa} et~al.}{2018}]{Correa_2018}
{Correa} C.~A.,  {Schaye} J.,  {Wyithe} J. S.~B.,  {Duffy} A.~R.,  {Theuns} T.,  {Crain} R.~A.,   {Bower} R.~G.,  2018, \mn@doi [\mnras] {10.1093/mnras/stx2332}, \href {https://ui.adsabs.harvard.edu/abs/2018MNRAS.473..538C} {473, 538}

\bibitem[\protect\citeauthoryear{{Crain} \& {van de Voort}}{{Crain} \& {van de Voort}}{2023}]{Crain2023}
{Crain} R.~A.,  {van de Voort} F.,  2023, \mn@doi [\araa] {10.1146/annurev-astro-041923-043618}, \href {https://ui.adsabs.harvard.edu/abs/2023ARA&A..61..473C} {61, 473}

\bibitem[\protect\citeauthoryear{{Crain} et~al.,}{{Crain} et~al.}{2015}]{Crain2015}
{Crain} R.~A.,  et~al., 2015, \mn@doi [\mnras] {10.1093/mnras/stv725}, \href {https://ui.adsabs.harvard.edu/abs/2015MNRAS.450.1937C} {450, 1937}

\bibitem[\protect\citeauthoryear{{Dai} \& {Seljak}}{{Dai} \& {Seljak}}{2021}]{Dai2021}
{Dai} B.,  {Seljak} U.,  2021, \mn@doi [Proceedings of the National Academy of Science] {10.1073/pnas.2020324118}, \href {https://ui.adsabs.harvard.edu/abs/2021PNAS..11820324D} {118, e2020324118}

\bibitem[\protect\citeauthoryear{{Decataldo}, {Shen}, {Mayer}, {Baumschlager}  \& {Madau}}{{Decataldo} et~al.}{2024}]{Decataldo2024}
{Decataldo} D.,  {Shen} S.,  {Mayer} L.,  {Baumschlager} B.,   {Madau} P.,  2024, \mn@doi [\aap] {10.1051/0004-6361/202346972}, \href {https://ui.adsabs.harvard.edu/abs/2024A&A...685A...8D} {685, A8}

\bibitem[\protect\citeauthoryear{{Dekel} \& {Birnboim}}{{Dekel} \& {Birnboim}}{2006}]{Dekel2006}
{Dekel} A.,  {Birnboim} Y.,  2006, \mn@doi [\mnras] {10.1111/j.1365-2966.2006.10145.x}, \href {https://ui.adsabs.harvard.edu/abs/2006MNRAS.368....2D} {368, 2}

\bibitem[\protect\citeauthoryear{{Delgado} et~al.,}{{Delgado} et~al.}{2023}]{Delgado2023}
{Delgado} A.~M.,  et~al., 2023, \mn@doi [\mnras] {10.1093/mnras/stad2992}, \href {https://ui.adsabs.harvard.edu/abs/2023MNRAS.526.5306D} {526, 5306}

\bibitem[\protect\citeauthoryear{{Diemand}, {Kuhlen}  \& {Madau}}{{Diemand} et~al.}{2007}]{Diemand2007}
{Diemand} J.,  {Kuhlen} M.,   {Madau} P.,  2007, \mn@doi [\apj] {10.1086/520573}, \href {https://ui.adsabs.harvard.edu/abs/2007ApJ...667..859D} {667, 859}

\bibitem[\protect\citeauthoryear{{Donnari} et~al.,}{{Donnari} et~al.}{2021}]{Donnari2021}
{Donnari} M.,  et~al., 2021, \mn@doi [\mnras] {10.1093/mnras/staa3006}, \href {https://ui.adsabs.harvard.edu/abs/2021MNRAS.500.4004D} {500, 4004}

\bibitem[\protect\citeauthoryear{{Dubois}, {Gavazzi}, {Peirani}  \& {Silk}}{{Dubois} et~al.}{2013}]{Dubois2013}
{Dubois} Y.,  {Gavazzi} R.,  {Peirani} S.,   {Silk} J.,  2013, \mn@doi [\mnras] {10.1093/mnras/stt997}, \href {https://ui.adsabs.harvard.edu/abs/2013MNRAS.433.3297D} {433, 3297}

\bibitem[\protect\citeauthoryear{{Dubois}, {Peirani}, {Pichon}, {Devriendt}, {Gavazzi}, {Welker}  \& {Volonteri}}{{Dubois} et~al.}{2016}]{Dubois2016}
{Dubois} Y.,  {Peirani} S.,  {Pichon} C.,  {Devriendt} J.,  {Gavazzi} R.,  {Welker} C.,   {Volonteri} M.,  2016, \mn@doi [\mnras] {10.1093/mnras/stw2265}, \href {https://ui.adsabs.harvard.edu/abs/2016MNRAS.463.3948D} {463, 3948}

\bibitem[\protect\citeauthoryear{Dzanic, Shah  \& Witherden}{Dzanic et~al.}{2019}]{Dzanic2019}
Dzanic T.,  Shah K.,   Witherden F.,  2019 (\mn@eprint {arXiv} {1911.06465})

\bibitem[\protect\citeauthoryear{{Faucher-Gigu{\`e}re}, {Kere{\v{s}}}  \& {Ma}}{{Faucher-Gigu{\`e}re} et~al.}{2011}]{Faucher2011}
{Faucher-Gigu{\`e}re} C.-A.,  {Kere{\v{s}}} D.,   {Ma} C.-P.,  2011, \mn@doi [\mnras] {10.1111/j.1365-2966.2011.19457.x}, \href {https://ui.adsabs.harvard.edu/abs/2011MNRAS.417.2982F} {417, 2982}

\bibitem[\protect\citeauthoryear{{Faucher-Giguère} \& {Oh}}{{Faucher-Giguère} \& {Oh}}{2023}]{Faucher_2023}
{Faucher-Giguère} C.-A.,  {Oh} S.~P.,  2023, \mn@doi [\araa] {10.1146/annurev-astro-052920-125203}, \href {https://ui.adsabs.harvard.edu/abs/2023ARA&A..61..131F} {61, 131}

\bibitem[\protect\citeauthoryear{{Faucher-Giguère}, {Hopkins}, {Kere{\v{s}}}, {Muratov}, {Quataert}  \& {Murray}}{{Faucher-Giguère} et~al.}{2015}]{Faucher2015}
{Faucher-Giguère} C.-A.,  {Hopkins} P.~F.,  {Kere{\v{s}}} D.,  {Muratov} A.~L.,  {Quataert} E.,   {Murray} N.,  2015, \mn@doi [\mnras] {10.1093/mnras/stv336}, \href {https://ui.adsabs.harvard.edu/abs/2015MNRAS.449..987F} {449, 987}

\bibitem[\protect\citeauthoryear{{Faucher-Giguère}, {Feldmann}, {Quataert}, {Kere{\v{s}}}, {Hopkins}  \& {Murray}}{{Faucher-Giguère} et~al.}{2016}]{Faucher2016}
{Faucher-Giguère} C.-A.,  {Feldmann} R.,  {Quataert} E.,  {Kere{\v{s}}} D.,  {Hopkins} P.~F.,   {Murray} N.,  2016, \mn@doi [\mnras] {10.1093/mnrasl/slw091}, \href {https://ui.adsabs.harvard.edu/abs/2016MNRAS.461L..32F} {461, L32}

\bibitem[\protect\citeauthoryear{{Feldmann} et~al.,}{{Feldmann} et~al.}{2023}]{Feldmann2023}
{Feldmann} R.,  et~al., 2023, \mn@doi [\mnras] {10.1093/mnras/stad1205}, \href {https://ui.adsabs.harvard.edu/abs/2023MNRAS.522.3831F} {522, 3831}

\bibitem[\protect\citeauthoryear{Fuoli, Van~Gool  \& Timofte}{Fuoli et~al.}{2021}]{Fuoli2021}
Fuoli D.,  Van~Gool L.,   Timofte R.,  2021 (\mn@eprint {arXiv} {2106.00783})

\bibitem[\protect\citeauthoryear{{Gebhardt} et~al.,}{{Gebhardt} et~al.}{2024}]{Gebhardt2024}
{Gebhardt} M.,  et~al., 2024, \mn@doi [\mnras] {10.1093/mnras/stae817}, \href {https://ui.adsabs.harvard.edu/abs/2024MNRAS.529.4896G} {529, 4896}

\bibitem[\protect\citeauthoryear{{Genel} et~al.,}{{Genel} et~al.}{2019}]{Genel2019}
{Genel} S.,  et~al., 2019, \mn@doi [\apj] {10.3847/1538-4357/aaf4bb}, \href {https://ui.adsabs.harvard.edu/abs/2019ApJ...871...21G} {871, 21}

\bibitem[\protect\citeauthoryear{{Gensior}, {Feldmann}, {Mayer}, {Wetzel}, {Hopkins}  \& {Faucher-Gigu{\`e}re}}{{Gensior} et~al.}{2023}]{Gensior2023}
{Gensior} J.,  {Feldmann} R.,  {Mayer} L.,  {Wetzel} A.,  {Hopkins} P.~F.,   {Faucher-Gigu{\`e}re} C.-A.,  2023, \mn@doi [\mnras] {10.1093/mnrasl/slac138}, \href {https://ui.adsabs.harvard.edu/abs/2023MNRAS.518L..63G} {518, L63}

\bibitem[\protect\citeauthoryear{{Gensior} et~al.,}{{Gensior} et~al.}{2024}]{Gensior2024}
{Gensior} J.,  et~al., 2024, \mn@doi [\mnras] {10.1093/mnras/stae1217}, \href {https://ui.adsabs.harvard.edu/abs/2024MNRAS.tmp.1249G} {}

\bibitem[\protect\citeauthoryear{Goodfellow}{Goodfellow}{2016}]{Goodfellow2016}
Goodfellow I.,  2016, NIPS 2016 Tutorial: Generative Adversarial Networks (\mn@eprint {arXiv} {1701.00160})

\bibitem[\protect\citeauthoryear{{Goodfellow}, {Pouget-Abadie}, {Mirza}, {Xu}, {Warde-Farley}, {Ozair}, {Courville}  \& {Bengio}}{{Goodfellow} et~al.}{2014}]{Goodfellow2014}
{Goodfellow} I.~J.,  {Pouget-Abadie} J.,  {Mirza} M.,  {Xu} B.,  {Warde-Farley} D.,  {Ozair} S.,  {Courville} A.,   {Bengio} Y.,  2014, {Generative Adversarial Networks} (\mn@eprint {arXiv} {1406.2661})

\bibitem[\protect\citeauthoryear{Gulrajani, Ahmed, Arjovsky, Dumoulin  \& Courville}{Gulrajani et~al.}{2017}]{Gulrajani2017}
Gulrajani I.,  Ahmed F.,  Arjovsky M.,  Dumoulin V.,   Courville A.,  2017, Improved Training of Wasserstein GANs (\mn@eprint {arXiv} {1704.00028})

\bibitem[\protect\citeauthoryear{{Hafen} et~al.,}{{Hafen} et~al.}{2019}]{Hafen2019}
{Hafen} Z.,  et~al., 2019, \mn@doi [\mnras] {10.1093/mnras/stz1773}, 488, 1248

\bibitem[\protect\citeauthoryear{{Hafen} et~al.,}{{Hafen} et~al.}{2020}]{Hafen2020}
{Hafen} Z.,  et~al., 2020, \mn@doi [\mnras] {10.1093/mnras/staa902}, 494, 3581

\bibitem[\protect\citeauthoryear{Hahn \& Abel}{Hahn \& Abel}{2011}]{Hahn2011}
Hahn O.,  Abel T.,  2011, \mn@doi [\mnras] {10.1111/j.1365-2966.2011.18820.x}, 415, 2101–2121

\bibitem[\protect\citeauthoryear{{Harrington}, {Mustafa}, {Dornfest}, {Horowitz}  \& {Luki{\'c}}}{{Harrington} et~al.}{2022}]{Harrington2022}
{Harrington} P.,  {Mustafa} M.,  {Dornfest} M.,  {Horowitz} B.,   {Luki{\'c}} Z.,  2022, \mn@doi [\apj] {10.3847/1538-4357/ac5faa}, \href {https://ui.adsabs.harvard.edu/abs/2022ApJ...929..160H} {929, 160}

\bibitem[\protect\citeauthoryear{{Hascoet}, {Febvre}, {Ariki}  \& {Takiguchi}}{{Hascoet} et~al.}{2019}]{Hascoet2019}
{Hascoet} T.,  {Febvre} Q.,  {Ariki} Y.,   {Takiguchi} T.,  2019 (\mn@eprint {arXiv} {1910.11127})

\bibitem[\protect\citeauthoryear{Hassan et~al.,}{Hassan et~al.}{2022}]{Hassan2022}
Hassan S.,  et~al., 2022, \mn@doi [\apj] {10.3847/1538-4357/ac8b09}, 937, 83

\bibitem[\protect\citeauthoryear{He, Zhang, Ren  \& Sun}{He et~al.}{2015}]{He2015}
He K.,  Zhang X.,  Ren S.,   Sun J.,  2015 (\mn@eprint {arXiv} {1512.03385})

\bibitem[\protect\citeauthoryear{Heusel, Ramsauer, Unterthiner, Nessler  \& Hochreiter}{Heusel et~al.}{2017}]{Heusel2017}
Heusel M.,  Ramsauer H.,  Unterthiner T.,  Nessler B.,   Hochreiter S.,  2017 (\mn@eprint {arXiv} {1706.08500})

\bibitem[\protect\citeauthoryear{Ho, Martin  \& Turner}{Ho et~al.}{2019}]{Ho2019}
Ho S.~H.,  Martin C.~L.,   Turner M.~L.,  2019, \mn@doi [\apj] {10.3847/1538-4357/ab0ec2}, 875, 54

\bibitem[\protect\citeauthoryear{{Hoeft}, {M{\"u}cket}  \& {Gottl{\"o}ber}}{{Hoeft} et~al.}{2004}]{Hoeft2004}
{Hoeft} M.,  {M{\"u}cket} J.~P.,   {Gottl{\"o}ber} S.,  2004, \mn@doi [\apj] {10.1086/380990}, \href {https://ui.adsabs.harvard.edu/abs/2004ApJ...602..162H} {602, 162}

\bibitem[\protect\citeauthoryear{{Hopkins}}{{Hopkins}}{2015}]{Hopkins2015}
{Hopkins} P.~F.,  2015, \mn@doi [\mnras] {10.1093/mnras/stv195}, \href {https://ui.adsabs.harvard.edu/abs/2015MNRAS.450...53H} {450, 53}

\bibitem[\protect\citeauthoryear{Hopkins et~al.,}{Hopkins et~al.}{2018}]{Hopkins2018}
Hopkins P.~F.,  et~al., 2018, \mn@doi [\mnras] {10.1093/mnras/sty1690}, 480, 800–863

\bibitem[\protect\citeauthoryear{{Horowitz}, {Dornfest}, {Luki{\'c}}  \& {Harrington}}{{Horowitz} et~al.}{2022}]{Horowitz2022}
{Horowitz} B.,  {Dornfest} M.,  {Luki{\'c}} Z.,   {Harrington} P.,  2022, \mn@doi [\apj] {10.3847/1538-4357/ac9ea7}, \href {https://ui.adsabs.harvard.edu/abs/2022ApJ...941...42H} {941, 42}

\bibitem[\protect\citeauthoryear{Isola, Zhu, Zhou  \& Efros}{Isola et~al.}{2016}]{Isola2016}
Isola P.,  Zhu J.-Y.,  Zhou T.,   Efros A.~A.,  2016 (\mn@eprint {arXiv} {1611.07004})

\bibitem[\protect\citeauthoryear{{Jamieson}, {Li}, {de Oliveira}, {Villaescusa-Navarro}, {Ho}  \& {Spergel}}{{Jamieson} et~al.}{2023}]{Jamieson2023}
{Jamieson} D.,  {Li} Y.,  {de Oliveira} R.~A.,  {Villaescusa-Navarro} F.,  {Ho} S.,   {Spergel} D.~N.,  2023, \mn@doi [\apj] {10.3847/1538-4357/acdb6c}, \href {https://ui.adsabs.harvard.edu/abs/2023ApJ...952..145J} {952, 145}

\bibitem[\protect\citeauthoryear{Jung \& Keuper}{Jung \& Keuper}{2020}]{Jung2020}
Jung S.,  Keuper M.,  2020 (\mn@eprint {arXiv} {2012.03110})

\bibitem[\protect\citeauthoryear{Karnewar \& Wang}{Karnewar \& Wang}{2019}]{Karnewar2019}
Karnewar A.,  Wang O.,  2019, MSG-GAN: Multi-Scale Gradients for Generative Adversarial Networks (\mn@eprint {arXiv} {1903.06048})

\bibitem[\protect\citeauthoryear{Karras, Aila, Laine  \& Lehtinen}{Karras et~al.}{2017}]{Karras2017}
Karras T.,  Aila T.,  Laine S.,   Lehtinen J.,  2017 (\mn@eprint {arXiv} {1710.10196})

\bibitem[\protect\citeauthoryear{Karras, Laine  \& Aila}{Karras et~al.}{2018}]{Karras2018}
Karras T.,  Laine S.,   Aila T.,  2018 (\mn@eprint {arXiv} {1812.04948})

\bibitem[\protect\citeauthoryear{Kaushal, Villaescusa-Navarro, Giusarma, Li, Hawry  \& Reyes}{Kaushal et~al.}{2022}]{Kaushal2022}
Kaushal N.,  Villaescusa-Navarro F.,  Giusarma E.,  Li Y.,  Hawry C.,   Reyes M.,  2022, \mn@doi [\apj] {10.3847/1538-4357/ac5c4a}, 930, 115

\bibitem[\protect\citeauthoryear{{Keller}, {Wadsley}, {Wang}  \& {Kruijssen}}{{Keller} et~al.}{2019}]{Keller2019}
{Keller} B.~W.,  {Wadsley} J.~W.,  {Wang} L.,   {Kruijssen} J.~M.~D.,  2019, \mn@doi [\mnras] {10.1093/mnras/sty2859}, \href {https://ui.adsabs.harvard.edu/abs/2019MNRAS.482.2244K} {482, 2244}

\bibitem[\protect\citeauthoryear{{Kere{\v{s}}}, {Katz}, {Weinberg}  \& {Dav{\'e}}}{{Kere{\v{s}}} et~al.}{2005}]{Keres2005}
{Kere{\v{s}}} D.,  {Katz} N.,  {Weinberg} D.~H.,   {Dav{\'e}} R.,  2005, \mn@doi [\mnras] {10.1111/j.1365-2966.2005.09451.x}, \href {https://ui.adsabs.harvard.edu/abs/2005MNRAS.363....2K} {363, 2}

\bibitem[\protect\citeauthoryear{Khayatkhoei \& Elgammal}{Khayatkhoei \& Elgammal}{2020}]{Khayatkhoei2020}
Khayatkhoei M.,  Elgammal A.,  2020 (\mn@eprint {arXiv} {2010.01473})

\bibitem[\protect\citeauthoryear{Kingma \& Ba}{Kingma \& Ba}{2017}]{Kingma2017}
Kingma D.~P.,  Ba J.,  2017, Adam: A Method for Stochastic Optimization (\mn@eprint {arXiv} {1412.6980})

\bibitem[\protect\citeauthoryear{{Kodi Ramanah}, {Charnock}, {Villaescusa-Navarro}  \& {Wandelt}}{{Kodi Ramanah} et~al.}{2020}]{Ramanah2020}
{Kodi Ramanah} D.,  {Charnock} T.,  {Villaescusa-Navarro} F.,   {Wandelt} B.~D.,  2020, \mn@doi [\mnras] {10.1093/mnras/staa1428}, \href {https://ui.adsabs.harvard.edu/abs/2020MNRAS.495.4227K} {495, 4227}

\bibitem[\protect\citeauthoryear{Li et~al.,}{Li et~al.}{2018}]{Li2018}
Li Y.-P.,  et~al., 2018, \mn@doi [\apj] {10.3847/1538-4357/aade8b}, 866, 70

\bibitem[\protect\citeauthoryear{{Li}, {Ni}, {Croft}, {Di Matteo}, {Bird}  \& {Feng}}{{Li} et~al.}{2021}]{Li2021}
{Li} Y.,  {Ni} Y.,  {Croft} R. A.~C.,  {Di Matteo} T.,  {Bird} S.,   {Feng} Y.,  2021, \mn@doi [Proceedings of the National Academy of Science] {10.1073/pnas.2022038118}, \href {https://ui.adsabs.harvard.edu/abs/2021PNAS..11822038L} {118, e2022038118}

\bibitem[\protect\citeauthoryear{{\L}okas}{{\L}okas}{2020}]{Lokas2020}
{\L}okas E.~L.,  2020, \mn@doi [Astronomy & Astrophysics] {10.1051/0004-6361/202037643}, 638, A133

\bibitem[\protect\citeauthoryear{{\L}okas, Majewski, Kazantzidis, Mayer, Carlin, Nidever  \& Moustakas}{{\L}okas et~al.}{2012}]{Lokas2012}
{\L}okas E.~L.,  Majewski S.~R.,  Kazantzidis S.,  Mayer L.,  Carlin J.~L.,  Nidever D.~L.,   Moustakas L.~A.,  2012, \mn@doi [\apj] {10.1088/0004-637x/751/1/61}, 751, 61

\bibitem[\protect\citeauthoryear{{Lovell}, {Wilkins}, {Thomas}, {Schaller}, {Baugh}, {Fabbian}  \& {Bah{\'e}}}{{Lovell} et~al.}{2022}]{Lovell2022}
{Lovell} C.~C.,  {Wilkins} S.~M.,  {Thomas} P.~A.,  {Schaller} M.,  {Baugh} C.~M.,  {Fabbian} G.,   {Bah{\'e}} Y.,  2022, \mn@doi [\mnras] {10.1093/mnras/stab3221}, \href {https://ui.adsabs.harvard.edu/abs/2022MNRAS.509.5046L} {509, 5046}

\bibitem[\protect\citeauthoryear{Maas}{Maas}{2013}]{Maas2013}
Maas A.~L.,  2013.

\bibitem[\protect\citeauthoryear{Mao, Liu, Shen, Li  \& Wang}{Mao et~al.}{2021}]{Mao2021}
Mao X.,  Liu Y.,  Shen W.,  Li Q.,   Wang Y.,  2021 (\mn@eprint {arXiv} {2111.11745})

\bibitem[\protect\citeauthoryear{Mathieu, Couprie  \& LeCun}{Mathieu et~al.}{2015}]{Mathieu2015}
Mathieu M.,  Couprie C.,   LeCun Y.,  2015 (\mn@eprint {arXiv} {1511.05440})

\bibitem[\protect\citeauthoryear{{Mori} \& {Burkert}}{{Mori} \& {Burkert}}{2000}]{Masao2000}
{Mori} M.,  {Burkert} A.,  2000, \mn@doi [\apj] {10.1086/309140}, \href {https://ui.adsabs.harvard.edu/abs/2000ApJ...538..559M} {538, 559}

\bibitem[\protect\citeauthoryear{{Nicola} et~al.,}{{Nicola} et~al.}{2022}]{Nicola2022}
{Nicola} A.,  et~al., 2022, \mn@doi [\jcap] {10.1088/1475-7516/2022/04/046}, \href {https://ui.adsabs.harvard.edu/abs/2022JCAP...04..046N} {2022, 046}

\bibitem[\protect\citeauthoryear{{Pandya} et~al.,}{{Pandya} et~al.}{2021}]{Pandya2021}
{Pandya} V.,  et~al., 2021, \mn@doi [\mnras] {10.1093/mnras/stab2714}, \href {https://ui.adsabs.harvard.edu/abs/2021MNRAS.508.2979P} {508, 2979}

\bibitem[\protect\citeauthoryear{Rahmati, Schaye, Pawlik  \& Raicevic}{Rahmati et~al.}{2013}]{Rahmati2013}
Rahmati A.,  Schaye J.,  Pawlik A.~H.,   Raicevic M.,  2013, \mn@doi [\mnras] {10.1093/mnras/stt324}, 431, 2261–2277

\bibitem[\protect\citeauthoryear{{Roca-Fabrega} et~al.,}{{Roca-Fabrega} et~al.}{2021}]{Roca2021}
{Roca-Fabrega} S.,  et~al., 2021, \mn@doi [\apj] {10.3847/1538-4357/ac088a}, \href {https://ui.adsabs.harvard.edu/abs/2021ApJ...917...64R} {917, 64}

\bibitem[\protect\citeauthoryear{{Rodriguez-Gomez} et~al.,}{{Rodriguez-Gomez} et~al.}{2019}]{Rodriguez-Gomez2019}
{Rodriguez-Gomez} V.,  et~al., 2019, \mn@doi [\mnras] {10.1093/mnras/sty3345}, \href {https://ui.adsabs.harvard.edu/abs/2019MNRAS.483.4140R} {483, 4140}

\bibitem[\protect\citeauthoryear{{Rohr} et~al.,}{{Rohr} et~al.}{2022}]{Rohr2022}
{Rohr} E.,  et~al., 2022, \mn@doi [\mnras] {10.1093/mnras/stab3625}, \href {https://ui.adsabs.harvard.edu/abs/2022MNRAS.510.3967R} {510, 3967}

\bibitem[\protect\citeauthoryear{Ronneberger, Fischer  \& Brox}{Ronneberger et~al.}{2015}]{Ronneberger2015}
Ronneberger O.,  Fischer P.,   Brox T.,  2015, U-Net: Convolutional Networks for Biomedical Image Segmentation (\mn@eprint {arXiv} {1505.04597})

\bibitem[\protect\citeauthoryear{{S{\'a}nchez Almeida}, {Elmegreen}, {Mu{\~n}oz-Tu{\~n}{\'o}n}  \& {Elmegreen}}{{S{\'a}nchez Almeida} et~al.}{2014}]{Sanchez-Almeida2014}
{S{\'a}nchez Almeida} J.,  {Elmegreen} B.~G.,  {Mu{\~n}oz-Tu{\~n}{\'o}n} C.,   {Elmegreen} D.~M.,  2014, \mn@doi [\aapr] {10.1007/s00159-014-0071-1}, \href {https://ui.adsabs.harvard.edu/abs/2014A&ARv..22...71S} {22, 71}

\bibitem[\protect\citeauthoryear{{Schneider} \& {Teyssier}}{{Schneider} \& {Teyssier}}{2015}]{Schneider2015}
{Schneider} A.,  {Teyssier} R.,  2015, \mn@doi [\jcap] {10.1088/1475-7516/2015/12/049}, \href {https://ui.adsabs.harvard.edu/abs/2015JCAP...12..049S} {2015, 049}

\bibitem[\protect\citeauthoryear{Schwarz, Liao  \& Geiger}{Schwarz et~al.}{2021}]{Schwarz2021}
Schwarz K.,  Liao Y.,   Geiger A.,  2021 (\mn@eprint {arXiv} {2111.02447})

\bibitem[\protect\citeauthoryear{{Sharma}, {Dai}, {Villaescusa-Navarro}  \& {Seljak}}{{Sharma} et~al.}{2024}]{Sharma2024}
{Sharma} D.,  {Dai} B.,  {Villaescusa-Navarro} F.,   {Seljak} U.,  2024, \mn@doi [arXiv e-prints] {10.48550/arXiv.2401.15891}, \href {https://ui.adsabs.harvard.edu/abs/2024arXiv240115891S} {p. arXiv:2401.15891}

\bibitem[\protect\citeauthoryear{{Snyder} et~al.,}{{Snyder} et~al.}{2015}]{Snyder2015}
{Snyder} G.~F.,  et~al., 2015, \mn@doi [\mnras] {10.1093/mnras/stv2078}, \href {https://ui.adsabs.harvard.edu/abs/2015MNRAS.454.1886S} {454, 1886}

\bibitem[\protect\citeauthoryear{{Stern}, {Fielding}, {Faucher-Giguère}  \& {Quataert}}{{Stern} et~al.}{2020}]{Stern2020}
{Stern} J.,  {Fielding} D.,  {Faucher-Giguère} C.-A.,   {Quataert} E.,  2020, \mn@doi [\mnras] {10.1093/mnras/staa198}, \href {https://ui.adsabs.harvard.edu/abs/2020MNRAS.492.6042S} {492, 6042}

\bibitem[\protect\citeauthoryear{{Stiskalek}, {Bartlett}, {Desmond}  \& {Anbajagane}}{{Stiskalek} et~al.}{2022}]{Stiskalek2022}
{Stiskalek} R.,  {Bartlett} D.~J.,  {Desmond} H.,   {Anbajagane} D.,  2022, \mn@doi [\mnras] {10.1093/mnras/stac1609}, \href {https://ui.adsabs.harvard.edu/abs/2022MNRAS.514.4026S} {514, 4026}

\bibitem[\protect\citeauthoryear{{Strawn} et~al.,}{{Strawn} et~al.}{2024}]{Strawn2024}
{Strawn} C.,  et~al., 2024, \mn@doi [\apj] {10.3847/1538-4357/ad12cb}, \href {https://ui.adsabs.harvard.edu/abs/2024ApJ...962...29S} {962, 29}

\bibitem[\protect\citeauthoryear{{Thiele}, {Villaescusa-Navarro}, {Spergel}, {Nelson}  \& {Pillepich}}{{Thiele} et~al.}{2020}]{Thiele2020}
{Thiele} L.,  {Villaescusa-Navarro} F.,  {Spergel} D.~N.,  {Nelson} D.,   {Pillepich} A.,  2020, \mn@doi [\apj] {10.3847/1538-4357/abb80f}, \href {https://ui.adsabs.harvard.edu/abs/2020ApJ...902..129T} {902, 129}

\bibitem[\protect\citeauthoryear{{Tortora}, {Feldmann}, {Bernardini}  \& {Faucher-Gigu{\`e}re}}{{Tortora} et~al.}{2023}]{Tortora2023}
{Tortora} L.,  {Feldmann} R.,  {Bernardini} M.,   {Faucher-Gigu{\`e}re} C.-A.,  2023, {The HI covering fraction of Lyman Limit Systems in FIRE haloes} (\mn@eprint {} {2311.18000})

\bibitem[\protect\citeauthoryear{{Tr{\"o}ster}, {Ferguson}, {Harnois-D{\'e}raps}  \& {McCarthy}}{{Tr{\"o}ster} et~al.}{2019}]{Troester2019}
{Tr{\"o}ster} T.,  {Ferguson} C.,  {Harnois-D{\'e}raps} J.,   {McCarthy} I.~G.,  2019, \mn@doi [\mnras] {10.1093/mnrasl/slz075}, \href {https://ui.adsabs.harvard.edu/abs/2019MNRAS.487L..24T} {487, L24}

\bibitem[\protect\citeauthoryear{Valentini et~al.,}{Valentini et~al.}{2019}]{Valentini2019}
Valentini M.,  et~al., 2019, \mn@doi [\mnras] {10.1093/mnras/stz3131}

\bibitem[\protect\citeauthoryear{{Vaswani}, {Shazeer}, {Parmar}, {Uszkoreit}, {Jones}, {Gomez}, {Kaiser}  \& {Polosukhin}}{{Vaswani} et~al.}{2017}]{Vasvani_2017}
{Vaswani} A.,  {Shazeer} N.,  {Parmar} N.,  {Uszkoreit} J.,  {Jones} L.,  {Gomez} A.~N.,  {Kaiser} L.,   {Polosukhin} I.,  2017, {Attention Is All You Need} (\mn@eprint {arXiv} {1706.03762})

\bibitem[\protect\citeauthoryear{{Villaescusa-Navarro}, {Wandelt}, {Angl{\'e}s-Alc{\'a}zar}, {Genel}, {Zorrilla Mantilla}, {Ho}  \& {Spergel}}{{Villaescusa-Navarro} et~al.}{2020}]{Villaescusa2020b}
{Villaescusa-Navarro} F.,  {Wandelt} B.~D.,  {Angl{\'e}s-Alc{\'a}zar} D.,  {Genel} S.,  {Zorrilla Mantilla} J.~M.,  {Ho} S.,   {Spergel} D.~N.,  2020 (\mn@eprint {arXiv} {2011.05992})

\bibitem[\protect\citeauthoryear{{Villaescusa-Navarro} et~al.,}{{Villaescusa-Navarro} et~al.}{2021a}]{Villaescusa2021}
{Villaescusa-Navarro} F.,  et~al., 2021a, {Multifield Cosmology with Artificial Intelligence} (\mn@eprint {arXiv} {2109.09747})

\bibitem[\protect\citeauthoryear{{Villaescusa-Navarro} et~al.,}{{Villaescusa-Navarro} et~al.}{2021b}]{Villaescusa2021b}
{Villaescusa-Navarro} F.,  et~al., 2021b, \mn@doi [\apj] {10.3847/1538-4357/abf7ba}, \href {https://ui.adsabs.harvard.edu/abs/2021ApJ...915...71V} {915, 71}

\bibitem[\protect\citeauthoryear{{Vogelsberger}, {Marinacci}, {Torrey}  \& {Puchwein}}{{Vogelsberger} et~al.}{2020}]{Vogelsberger2020}
{Vogelsberger} M.,  {Marinacci} F.,  {Torrey} P.,   {Puchwein} E.,  2020, \mn@doi [Nature Reviews Physics] {10.1038/s42254-019-0127-2}, \href {https://ui.adsabs.harvard.edu/abs/2020NatRP...2...42V} {2, 42}

\bibitem[\protect\citeauthoryear{{Wadekar}, {Villaescusa-Navarro}, {Ho}  \& {Perreault-Levasseur}}{{Wadekar} et~al.}{2021}]{Wadekar2021}
{Wadekar} D.,  {Villaescusa-Navarro} F.,  {Ho} S.,   {Perreault-Levasseur} L.,  2021, \mn@doi [\apj] {10.3847/1538-4357/ac033a}, \href {https://ui.adsabs.harvard.edu/abs/2021ApJ...916...42W} {916, 42}

\bibitem[\protect\citeauthoryear{Wang \& Makarenko}{Wang \& Makarenko}{2021}]{Wang2021}
Wang Q.,  Makarenko M.,  2021 (\mn@eprint {arXiv} {2110.02873})

\bibitem[\protect\citeauthoryear{{Wang} et~al.,}{{Wang} et~al.}{2017}]{Wang2017}
{Wang} M.-Y.,  et~al., 2017, in American Astronomical Society Meeting Abstracts \#229. p. 416.03

\bibitem[\protect\citeauthoryear{{Wang}, {Xu}, {Gao}, {Guo}, {Qu}  \& {Pan}}{{Wang} et~al.}{2019}]{Wang2019}
{Wang} L.,  {Xu} D.,  {Gao} L.,  {Guo} Q.,  {Qu} Y.,   {Pan} J.,  2019, \mn@doi [\mnras] {10.1093/mnras/stz529}, \href {https://ui.adsabs.harvard.edu/abs/2019MNRAS.485.2083W} {485, 2083}

\bibitem[\protect\citeauthoryear{{Wang}, {Xu}, {Lu}, {Cai}, {Xiang}, {Mao}, {Springel}  \& {Hernquist}}{{Wang} et~al.}{2022}]{Wang2022}
{Wang} S.,  {Xu} D.,  {Lu} S.,  {Cai} Z.,  {Xiang} M.,  {Mao} S.,  {Springel} V.,   {Hernquist} L.,  2022, \mn@doi [\mnras] {10.1093/mnras/stab3167}, \href {https://ui.adsabs.harvard.edu/abs/2022MNRAS.509.3148W} {509, 3148}

\bibitem[\protect\citeauthoryear{{Wellons} et~al.,}{{Wellons} et~al.}{2023}]{Wellons2023}
{Wellons} S.,  et~al., 2023, \mn@doi [\mnras] {10.1093/mnras/stad511}, \href {https://ui.adsabs.harvard.edu/abs/2023MNRAS.520.5394W} {520, 5394}

\bibitem[\protect\citeauthoryear{Woods, Wadsley, Couchman, Stinson  \& Shen}{Woods et~al.}{2014}]{Woods2014}
Woods R.~M.,  Wadsley J.,  Couchman H. M.~P.,  Stinson G.,   Shen S.,  2014, \mn@doi [\mnras] {10.1093/mnras/stu895}, 442, 732–740

\bibitem[\protect\citeauthoryear{{Wright}, {Somerville}, {Lagos}, {Schaller}, {Dav{\'e}}, {Angl{\'e}s-Alc{\'a}zar}  \& {Genel}}{{Wright} et~al.}{2024}]{Wright2024}
{Wright} R.~J.,  {Somerville} R.~S.,  {Lagos} C. d.~P.,  {Schaller} M.,  {Dav{\'e}} R.,  {Angl{\'e}s-Alc{\'a}zar} D.,   {Genel} S.,  2024, \mn@doi [\mnras] {10.1093/mnras/stae1688}, \href {https://ui.adsabs.harvard.edu/abs/2024MNRAS.532.3417W} {532, 3417}

\bibitem[\protect\citeauthoryear{{Xu}, {Luo}, {Kang}, {Li}, {Li}, {Wang}  \& {Libeskind}}{{Xu} et~al.}{2022}]{Xu2022}
{Xu} Y.,  {Luo} Y.,  {Kang} X.,  {Li} Z.,  {Li} Z.,  {Wang} P.,   {Libeskind} N.,  2022, \mn@doi [\apj] {10.3847/1538-4357/ac53ab}, \href {https://ui.adsabs.harvard.edu/abs/2022ApJ...928..100X} {928, 100}

\bibitem[\protect\citeauthoryear{Yang, Hong, Jang, Zhao  \& Lee}{Yang et~al.}{2019}]{Yang2019}
Yang D.,  Hong S.,  Jang Y.,  Zhao T.,   Lee H.,  2019 (\mn@eprint {arXiv} {1901.09024})

\bibitem[\protect\citeauthoryear{{Zamudio-Fernandez}, {Okan}, {Villaescusa-Navarro}, {Bilaloglu}, {Derin Cengiz}, {He}, {Perreault Levasseur}  \& {Ho}}{{Zamudio-Fernandez} et~al.}{2019}]{Zamudio2019}
{Zamudio-Fernandez} J.,  {Okan} A.,  {Villaescusa-Navarro} F.,  {Bilaloglu} S.,  {Derin Cengiz} A.,  {He} S.,  {Perreault Levasseur} L.,   {Ho} S.,  2019 (\mn@eprint {arXiv} {1904.12846})

\bibitem[\protect\citeauthoryear{{Zhang}, {Goodfellow}, {Metaxas}  \& {Odena}}{{Zhang} et~al.}{2018}]{Zhang2_2018}
{Zhang} H.,  {Goodfellow} I.,  {Metaxas} D.,   {Odena} A.,  2018, {Self-Attention Generative Adversarial Networks} (\mn@eprint {arXiv} {1805.08318})

\bibitem[\protect\citeauthoryear{{Zhang}, {Lachance}, {Ni}, {Li}, {Croft}, {Matteo}, {Bird}  \& {Feng}}{{Zhang} et~al.}{2024}]{Zhang2024}
{Zhang} X.,  {Lachance} P.,  {Ni} Y.,  {Li} Y.,  {Croft} R. A.~C.,  {Matteo} T.~D.,  {Bird} S.,   {Feng} Y.,  2024, \mn@doi [\mnras] {10.1093/mnras/stad3940}, \href {https://ui.adsabs.harvard.edu/abs/2024MNRAS.528..281Z} {528, 281}

\bibitem[\protect\citeauthoryear{Zhou, Yu, Huang, Zhao, Gu, Loy, Meng  \& Li}{Zhou et~al.}{2022}]{Zhou2022}
Zhou M.,  Yu H.,  Huang J.,  Zhao F.,  Gu J.,  Loy C.~C.,  Meng D.,   Li C.,  2022 (\mn@eprint {arXiv} {2210.05171})

\bibitem[\protect\citeauthoryear{Zhu, Zhang, Pathak, Darrell, Efros, Wang  \& Shechtman}{Zhu et~al.}{2017}]{Zhu2017}
Zhu J.-Y.,  Zhang R.,  Pathak D.,  Darrell T.,  Efros A.~A.,  Wang O.,   Shechtman E.,  2017

\bibitem[\protect\citeauthoryear{van Daalen, McCarthy  \& Schaye}{van Daalen et~al.}{2019}]{vanDaalen2019}
van Daalen M.~P.,  McCarthy I.~G.,   Schaye J.,  2019, \mn@doi [\mnras] {10.1093/mnras/stz3199}, 491, 2424–2446

\bibitem[\protect\citeauthoryear{{van de Voort}, {Schaye}, {Altay}  \& {Theuns}}{{van de Voort} et~al.}{2012}]{vanDenVoort2015}
{van de Voort} F.,  {Schaye} J.,  {Altay} G.,   {Theuns} T.,  2012, \mn@doi [\mnras] {10.1111/j.1365-2966.2012.20487.x}, \href {https://ui.adsabs.harvard.edu/abs/2012MNRAS.421.2809V} {421, 2809}

\makeatother
\end{thebibliography}


\appendix
\section{Testing on an independent simulation}\label{app:test_independent_simulation}
\begin{figure}
    \includegraphics[width=\columnwidth]{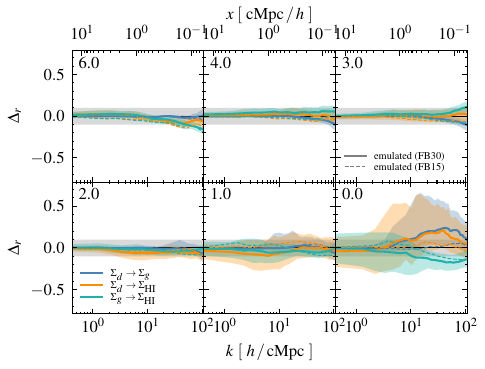}
    \caption{
    Cross-correlation errors $\Delta_r$ across scales and redshifts (indicated in the upper left corners) between the emulated fields from \embertwo{} and simulated fields. Shown are three cross-correlations between the input dark matter field and the output gas and HI surface densities, as well as the cross-correlation between the gas and HI surface density fields in green.
    The solid and dashed curves indicate whether the FB30 or FB15 input was used in the emulation.
    For the FB30 case, the shaded area and solid lines represent the median and 16th to 84th percentile at each $k$-scale, while the grey band indicates the 10 per cent error.
    For better visibility, we do not show additional percentiles for the FB15 case.}
    \label{fig:cross_corr_dm_256}
\end{figure}
\begin{figure}
    \includegraphics[width=\columnwidth]{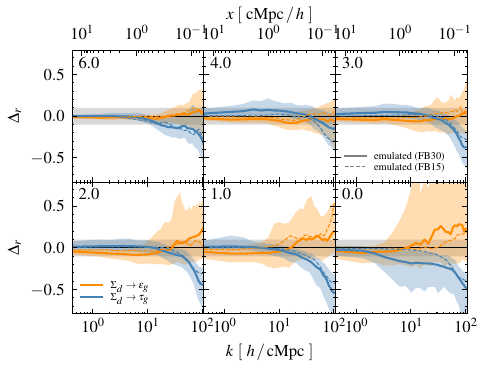}
    \caption{
    Cross-correlation errors $\Delta_r$ across scales and redshifts (indicated in the upper left corners) between the emulated fields from \embertwo{} and simulated fields. Shown are the cross-correlation errors for the kinetic and thermal surface energy densities $\varepsilon_g$ and $\tau_g$.
    The solid and dashed curves indicate whether the FB30 or FB15 input was used in the emulation.
    The shaded area and solid lines represent the median and 16th to 84th percentile at each $k$-scale, while the grey band indicates the 10 per cent error.
    For better visibility, we do not show additional percentiles for the dmo case.}
    \label{fig:cross_corr_derived_fields_256}
\end{figure}
The \embertwo{} model was trained with a cross-validation strategy, where we used two projections for training and the remaining for testing.
Since a given halo will appear from different angles in the training and test sets of \fbb{}, different projections can be, in principle, correlated and the networks prediction may be biased.
Here, we validate the models accuracy beyond the \fbb{} volume by reconstructing the target fields for an independent test set consisting of a simulation of a 15 cMpc$/h$ box, with the same physics and resolution as \fbb{}.
The results of this analysis are shown in figures \ref{fig:cross_corr_dm_256} and \ref{fig:cross_corr_derived_fields_256}, where we compare the emulation predictions between the \fbb{} and FB15.
We find that the models accuracy is independent on the test set, and thus conclude that the model has successfully learned to capture the relevant correlations.
\section{Implementation details}\label{app:additional_information}
We use \texttt{pytorch}\footnote{\url{https://pytorch.org}} for the implementation and training of the NNs. We use a fixed batch size of 4 and the Adam optimizer \citep[][]{Kingma2017} with $\beta_1=0.5$, $\beta_2=0.9$ and learning rate of $10^{-4}$.
The loss functions we used for the generator and discriminator are the ones presented in equation \ref{eq:L_G} and \ref{eq:L_D}, with no additional regularization techniques.
Similar to StyleGAN2 \citep[][]{Karras2018}, we adopt the Exponential-Moving Average (EMA) method to ensure model convergence. In EMA, the model parameters are updated using a weighted average of the current parameters and the previous EMA parameters. This helps stabilize the training process by smoothing out the fluctuations that can occur during optimization. The EMA update rule can be expressed mathematically as:
\begin{equation}
    \theta_{\text{EMA}} \leftarrow \alpha \cdot \theta_{\text{EMA}} + (1 - \alpha) \cdot \theta_{\text{current}},
\end{equation}
where $\theta_{\text{EMA}}$ denotes the parameters of the EMA model, $\theta_{\text{current}}$ represents the current model parameters, and $\alpha$ is a decay factor fixed to 0.995.
Using EMA allows us to retain a model that has averaged over several training iterations, which is used as the fiducial model in this paper.
We train for $5\times 10^6$ iterations, i.e. at the end of training a single network has processed $20 \times 10^6$ samples.
The training was performed on a single Tesla V-100 GPU card and took exactly one week per model, due to wallclock time limitations related to the local compute facilities.

\end{document}